\documentclass[final,acmsmall,screen,nonacm]{acmart}
\pdfoutput=1

\usepackage{ifthen}
\usepackage{enumitem}

\newcommand{\ifdraft}[2]{#2}

\newboolean{proofsinappendix}
\setboolean{proofsinappendix}{true}
\ifdraft{}{%
  \setboolean{proofsinappendix}{true}
}

\newboolean{dropappendix}
\setboolean{dropappendix}{true}

\newboolean{preprintversion}
\makeatletter
\@ifclasswith{acmart}{nonacm}{
  \setboolean{preprintversion}{true}
}{
  \setboolean{preprintversion}{false}
}
\makeatother
\ifthenelse{\boolean{preprintversion}}{%
  \setboolean{dropappendix}{false}
}{}

\usepackage[utf8]{inputenc}
\usepackage[T1]{fontenc}

\usepackage{siunitx}
\usepackage{makecell, booktabs, caption}
\usepackage{multirow}

\usepackage{bbm}
\usepackage{xspace}
\usepackage{newfile}
\usepackage{tikz}
\usepackage{algorithm}
\usepackage{algpseudocodex}
\usepackage{listings}
\usepackage{tabularx}
\usepackage{cleveref}

\ifdraft{
  \usepackage{todonotes}
}{
  \usepackage[disable]{todonotes}
}

\usepackage{twshowkeys}

\definecolor{docStyleBlue}{HTML}{337AB7}
\definecolor{docStyleGray}{HTML}{7A7A7A}

\setuptodonotes{
  backgroundcolor=yellow!30,
  size=\tiny,
  noline,
}

\newcommand{\jj}[1]{{\color{orange}\textbf{JJ:}~#1}}
\newcommand{\tw}[1]{{\color{green}\textbf{TW:}~#1}}
\newcommand{\jjnote}[1]{\todo[backgroundcolor=blue!50!white]{JJ:~#1}}
\newcommand{\twnote}[1]{\todo[backgroundcolor=green!50!white]{TW:~#1}}
\newcommand{\citetodo}[1]{[{\color{red!40!black}#1}]}

\ifdraft{}{%
\renewcommand{\jj}[1]{}
\renewcommand{\tw}[1]{}
\renewcommand{\jjnote}[1]{}
\renewcommand{\twnote}[1]{}
\renewcommand{\citetodo}[1]{}
}

\renewcommand{\gets}{\ensuremath{\xspace\mathop{:=}}}

\makeatletter
\ifthenelse{\boolean{proofsinappendix}}{
  \newoutputstream{proofstream}
  \openoutputfile{\jobname-proofs.out}{proofstream}
}{
}

\newcommand{\proofappendixbegin}[2]{%
  \phantomsection%
  \subsection*{\textbf{#1~\Cref{#2}}}%
  \addcontentsline{toc}{subsection}{#1~\Cref{#2}}%
  \label{#2:proof}%
  \def\proofappendix@qedsymbolmissing{\qed}
}
\newcommand{\proofappendixend}{%
  \proofappendix@qedsymbolmissing%
}
\def\qedhere{\global\def\proofappendix@qedsymbolmissing{}\qed}

\newenvironment{proofhere}[2][Proof of]{%
  \subsection*{#1~\Cref{#2}.}%
  \def\proofappendix@qedsymbolmissing{\qed}%
}{%
  \proofappendixend%
}

\ifthenelse{\boolean{proofsinappendix}}{%
  \NewEnviron{proofappendix}[2][Proof of]{%
    \addtostream{proofstream}{\noexpand\proofappendixbegin{#1}{#2}}%
    \expandafter\addtostream{proofstream}{\expandonce{\BODY}}%
    \addtostream{proofstream}{\noexpand\proofappendixend}%
    \addtostream{proofstream}{}%
  }%
}{%
  \newenvironment{proofappendix}[2][Proof of]{%
    \begin{proofhere}[#1]{#2}
  }{%
    \end{proofhere}
  }
}
\makeatother

\newcommand{\resetCurThmBraces}{%
  \gdef\curThmBraceOpen{(}%
  \gdef\curThmBraceClose{)}}
\resetCurThmBraces
\newcommand{\removeThmBraces}{%
  \gdef\curThmBraceOpen{}%
  \gdef\curThmBraceClose{}}
\resetCurThmBraces

\newenvironment{notheorembrackets}{\removeThmBraces}{\resetCurThmBraces}

\usepackage{etoolbox}
\makeatletter
\patchcmd{\thmhead@acmdefinition}{(#3)}{\curThmBraceOpen #3\curThmBraceClose }{}{}
\patchcmd{\thmhead@acmplain}{(#3)}{\curThmBraceOpen #3\curThmBraceClose }{}{}
\makeatother

\newcommand{\closure}[1]{\ensuremath{\mathord{\uparrow}#1}}

\newcommand{\Q}{\ensuremath{\mathbb{Q}}\xspace}
\newcommand{\CO}{\ensuremath{\mathcal{O}}\xspace}
\newcommand{\CT}{\ensuremath{\Theta}\xspace}
\newcommand{\Z}{\ensuremath{\mathbb{Z}}\xspace}

\newcommand{\N}{\ensuremath{\mathbb{N}}\xspace}
\newcommand{\M}{\ensuremath{\mathcal{N}}\xspace}
\newcommand{\Set}{\ensuremath{\mathsf{Set}}\xspace}

\newcommand{\ar}{\ensuremath{\mathsf{ar}}}

\newcommand{\ceil}[1]{\ensuremath{\lceil #1 \rceil}}
\newcommand{\set}[2][]{\ensuremath{{#1\{#2#1\}}}}

\newcommand{\sig}{\ensuremath{\mathsf{sig}}}
\newcommand{\pred}{\ensuremath{\mathsf{pred}}}
\newcommand{\old}{\ensuremath{\mathsf{old}}}

\newcommand{\Powf}{\ensuremath{\mathcal{P}_{\mathsf{f}}}}

\newcommand{\etal}{\text{et al.}\xspace}

\newcommand{\Dist}{\ensuremath{\mathcal{D}}}
\newcommand{\renumber}{\ensuremath{\textsc{Renumber}}}
\newcommand{\Split}{\ensuremath{\textsc{Split}}}
\newcommand{\MarkDirty}{\ensuremath{\textsc{MarkDirty}}}
\newcommand{\atom}{\ensuremath{\mathsf{atom}}}
\newcommand{\dirty}{\ensuremath{\mathsf{di}}}
\newcommand{\clean}{\ensuremath{\mathsf{cl}}}

\newcommand{\textqt}[1]{\text{``}#1\text{''}}
\newcommand{\epito}{\ensuremath{\twoheadrightarrow}}
\newcommand{\monoto}{\ensuremath{\rightarrowtail}}

\def\argmax{\qopname\relax m{argmax}}

\newcommand{\sizeof}[1]{\ensuremath{\sharp(#1)}}
\renewcommand{\Im}{\ensuremath{\mathscr{I\!\!m}}}

\newcommand*{\eg}{\emph{e.g.,}\@\xspace}
\newcommand*{\ie}{\emph{i.e.,}\@\xspace}

\newcommand{\operation}[1]{\mathsf{#1}}
\newcommand{\member}[1]{\texttt{#1}}
\newcommand{\range}[2]{\ensuremath{#1..#2}}

\newcommand{\BCone}{B_{\clean_1}}
\newcommand{\BD}{B_{\dirty}}

\newcommand{\block}{\member{block\_of}}
\newcommand{\states}{\member{loc2state}}
\newcommand{\blocks}{\member{blocks}}
\newcommand{\loc}{\member{state2loc}}
\newcommand{\dirtyStates}{\ensuremath{dirty}}
\newcommand{\worklist}{\member{worklist}}
\newcommand{\imax}{\ensuremath{i_{max}}}

\newcommand{\AlgoSeparator}{} %
\tikzset{
  algo block brace node/.style={
    align=left,
    text=gray,
  },
  fixed width/.style={
  },
  algo block brace/.style={
    append after command={
      \pgfextra{
        \draw[decorate,decoration={brace,raise=2pt,amplitude=10pt},
        line width=1pt,
        line cap=round,
        draw=gray,
        brace shift/.style={
          xshift=9cm,
        },
        ]
        ([brace shift]\tikzlastnode.north west)
        --
        node[xshift=12pt,anchor=west,algo block brace node] {#1}
        ([brace shift]\tikzlastnode.south west);
      }
    }
  },
  algo block brace alt/.style={
    append after command={
      \pgfextra{
        \draw[decorate,decoration={brace,raise=2pt,amplitude=10pt},
        line width=1pt,
        line cap=round,
        draw=gray,
        brace shift/.style={
          xshift=7.5cm,
        },
        ]
        ([brace shift]\tikzlastnode.north west)
        --
        node[xshift=12pt,anchor=west,algo block brace node] {#1}
        ([brace shift]\tikzlastnode.south west);
      }
    }
  }
}

\usetikzlibrary{cd,calc,decorations.pathreplacing,calligraphy}

\newcommand{\thetool}{\textit{Boa}}

\begin{document}

\setlist[itemize]{leftmargin=*}
\setlist[enumerate]{leftmargin=*}

\title{Fast Coalgebraic Bisimilarity Minimization}

\author{Jules Jacobs}
\affiliation{
  \institution{Radboud University}
  \city{Nijmegen}
  \country{The Netherlands}
}
\author{Thorsten Wißmann}
\affiliation{
  \institution{Radboud University}
  \city{Nijmegen}
  \country{The Netherlands}
}



\begin{abstract}
  Coalgebraic bisimilarity minimization generalizes classical automaton
  minimization to a large class of automata whose transition structure is specified by a functor,
  subsuming strong, weighted, and probabilistic bisimilarity.
  This offers the enticing possibility of turning bisimilarity minimization into an off-the-shelf technology, without having to develop a new algorithm for each new type of automaton.
  Unfortunately, there is no existing algorithm that is fully general, efficient, and able to handle large systems.

  We present a generic algorithm that minimizes coalgebras over an arbitrary functor
  in the category of sets as long as the action on morphisms is sufficiently computable. The functor makes
  at most $\mathcal{O}(m \log n)$ calls to the functor-specific action,
  where $n$ is the number of states and $m$ is the number of transitions in the coalgebra.

  While more specialized algorithms can be asymptotically faster than our algorithm (usually by a factor of $\CO(\frac{m}{n})$),
  our algorithm is especially well suited to efficient implementation, and our tool \thetool{} often uses much less time and memory on existing benchmarks,
  and can handle larger automata, despite being more generic.
\end{abstract}

\maketitle

\ifthenelse{\boolean{preprintversion}}{\vfill}{}
\section{Introduction}
State-based systems arise in various shapes throughout computer science: as
automata for regular expressions,
as control-flow graphs of programs,
Markov decision processes,
(labelled) transition systems,
or as the small-step semantics of programming languages.
If the programming language of interest involves concurrency,
bisimulation can capture whether two systems exhibit the same behaviour~\cite{Winskel93,Milner1980}.
In model checking, a state-based system is derived from the implementation and
then checked against its specification.

It is often beneficial to reduce the size of a state-based system by merging all equivalent states.
Moore's algorithm \cite{Moore} and Hopcroft's $\CO(n \log n)$ algorithm \cite{Hopcroft71} do this for the deterministic finite automata that arise from regular expressions,
and produce the equivalent automaton with minimal number of states.
In model checking, state-space reduction can be effective as a preprocessing step~\cite{BaierKatoen08}.
For instance, in probabilistic model checking,
the time saved in model checking due to the smaller system exceeds the time needed to minimize the system~\cite{KatoenEA07}.

Subsequent to \citet{Hopcroft71},
a variety of algorithms were developed for minimizing different types of automata.
Examples are algorithms for
\ifthenelse{\boolean{preprintversion}}{\pagebreak[10]}{}
\begin{itemize}
\item transition systems (without action labels)~\cite{KanellakisSmolka83,KanellakisS90},
labelled transition systems~\cite{Valmari09}, which arise from the verification concurrent systems,
\item weighted bisimilarity~\cite{ValmariF10} for Markov chains and probabilistic settings (such as probabilistic model checking \cite{KatoenEA07}),
\item Markov decision processes~\cite{BaierEM00,GrooteEA18} that combine concurrency with probabilistic branching,
\item weighted tree automata~\cite{HoegbergEA09,HoegbergEA07} that arise in natural language processing~\cite{MayKnight06}.
\end{itemize}

Recently, those algorithms and system equivalences were subsumed by a coalgebraic
generalization~\cite{DorschEA17,coparFM19,WissmannEA2021}\tw{add citation to concurSpecialIssue in the final version}.
This generic algorithm is parametrized by a (Set-)functor that describes the concrete system
type of interest. Functors are a standard notion in category theory and a
key notion in the Haskell programming language.
In coalgebraic automaton minimization, the functor is used to attach transition data to each state of the automaton.
For instance, the powerset functor models non-deterministic branching in transition systems,
and the probability distribution functor models probabilistic branching in Markov chains.

The users of a coalgebraic minimization algorithm may create their own system type by composing the provided basic functors,
allowing them to freely combine deterministic, non-deterministic,
and probabilistic behaviour.
For instance, the functor to model Markov decision processes is the composition of the functors of transition systems
and the functor for probability distributions.
This generalization points to the enticing possibility of turning automata minimization for different types of automata into an off-the-shelf technology.

Unfortunately, there are two problems that currently block this vision.
Firstly, although the generic algorithm has excellent $\CO(m \log n)$ asymptotic complexity,
where $n$ is the number of states and $m$ is the number of edges,
it is slow in practice, and the data structures required for partition refinement suffer from hungry memory usage.
A machine with 16GB of RAM required several minutes to minimize tree automata with 150 thousand states and
ran out of memory when minimizing tree automata larger than 160 thousand states \cite{coparFM19,WissmannEA2021}.
This problem has also been observed for algorithms for specific automata types, e.g., transition systems~\cite{Valmari10}.
In order to increase the total memory available,
a distributed partition refinement algorithm has been developed \cite{BDM22},
(and previously also for specific automata types, e.g., labelled transition
systems~\cite{BlomOrzan05}), but this algorithm runs in $\CO(n^2)$ and requires expensive distributed hardware.

Secondly, the generic algorithm does not work for all Set-functors, because it places certain restrictions on the functor type necessary for the tight run time complexity.
For instance, the algorithm is not capable of minimizing frames for the monotone neighbourhood logic \cite{HansenKupke04,HansenKupke04cmcs}, arising in game theory~\cite{Parikh1985,Peleg87,Pauly2001}.

We present a new algorithm that works for \emph{all} system types given by computable $\Set{}$-functors,
requiring only an implementation of the functor's action on morphisms, which is
then used to compute so-called \emph{signatures of states}, a notion originally introduced for labelled transition systems~\cite{BlomOrzan05}.
The algorithm makes at most $\CO(m \log n$) calls to the functor implementation\tw{Maybe replace with: the algorithm computes at most ... state signatures},
where $n$ and $m$ are the number of states and edges in the automaton, respectively.
In almost all instances, one such call takes $\CO(k)$ time, where $k$ is the maximum out-degree of a state, so the overall run time is in $\CO(km\log n)$.
We compensate for this extra factor because our algorithm has been designed to be efficient in practice
and does not need large data structures: we only need the automaton with predecessors and a refinable partition data structure.

We provide an implementation of our algorithm in our tool called \thetool{}.
The user of the tool can either encode their system type as a composition of the functors natively supported by \thetool{},
or extend \thetool{} with a new functor by providing a small amount of Rust code that implements the functor's action on morphisms.

Empirical evaluation of our implementation shows that the memory usage is much reduced,
in certain cases by more than 100x compared to the distributed algorithm~\cite{BDM22},
such that the benchmarks that were used to illustrate its scalability can now be solved on a single computer.
Running time is also much reduced, in certain cases by more than 3000x, even though we run on a single core rather than a distributed cluster.
We believe that this is a major step towards coalgebraic partition refinement as an off-the-shelf technology for automaton minimization.

\paragraph{The rest of the paper is structured as follows.}

\begin{description}
  \item[\Cref{sec:CoalgebraIntro}:] Coalgebraic bisimilarity minimization and our algorithm in a nutshell.
  \item[\Cref{sec:formalcoalg}:] The formal statement of behavioural equivalence of states, and examples for how this reduces to known notions of equivalence for particular instantiations.
  \item[\Cref{sec:CoalgebraicPartitionRefinement}:] Detailed description of our coalgebraic minimization algorithm for any computable set functor, and time complexity analysis showing that the algorithm makes at most $\CO(m \log n)$ calls to the functor operation.
  \item[\Cref{sec:Instances}:] Instantiations of the algorithm showing its genericity.
  \item[\Cref{sec:Benchmarks}:] Benchmark results showing our algorithm outperforms earlier work.
  \item[\Cref{sec:Conclusion}:] Conclusion and future work.
\end{description}

\section{Fast Coalgebraic Bisimilarity Minimization in a Nutshell}
\label{sec:CoalgebraIntro}

\newcommand{\st}[1]{\mathbf{#1}}
\tikzstyle{state} = [thick, circle, draw=black, minimum size=0.3cm, scale=0.8]
\tikzstyle{accepting}=[double distance=1pt, outer sep=1pt+\pgflinewidth,inner sep=3pt]
\tikzstyle{arr} = [thick,->,>=stealth,shorten >= 1pt, shorten <= 1pt]
\renewcommand{\arraystretch}{1.2}
\newcommand{\FALSE}{\mathsf{F}}
\newcommand{\TRUE}{\mathsf{T}}

This section presents the key ideas of our fast coalgebraic minimization algorithm.
We start with an introduction to coalgebra,
and how the language of category theory provides an elegant unifying framework for different types of automata.
No knowledge of category theory is assumed;
we will go from the concrete to the abstract,
and category theoretic notions have been erased from the presentation as much as possible.

Let us thus start by looking at three examples of automata:
deterministic finite automata on the alphabet $\{a,b\}$,
transition systems,
and Markov chains.
The usual way of visualizing is depicted in the first row of \Cref{fig:coalgexamples}.
For instance, a deterministic finite automaton on state set $C$ is usually described via
a transition function $\delta \colon C \times \{a,b\} \to C$ and a set of accepting states $F \subseteq C$ (the initial state is not relevant for the task of computing equivalent states).
In order to generalize various types of automata, however, we take a \emph{state-centric} point of view,
where we consider all the data as being \emph{attached to a particular state}:
\begin{itemize}
  \item In a finite automaton on the alphabet $\{a,b\}$ each state has two successors: one for the input letter $a$ and one for the input letter $b$.
  Each state also carries a boolean that determines whether the state is accepting (double border), or not (single border).
  For instance, state $\st{3}$ in the deterministic automaton in the left column of \Cref{fig:coalgexamples} is not accepting, but after transitioning via $a$ it goes to state $\st{5}$, which is accepting.
  We can specify any deterministic automaton entirely via a map
  \[
    c\colon C\to \set{\FALSE,\TRUE}\times C\times C
  \]
  This map sends every state $q\in C$ to $(b,q_a,q_b) := c(q)$, where $b\in \set{\FALSE,\TRUE}$ specifies if $b$ is accepting, and $q_a, q_b\in C$ are the target states for in input $a$ and $b$, respectively. %

  \item A transition system consists of a (finite) set of locations $C$, plus a (finite) set of transitions $\textqt{\mathord{\to}} \subseteq C\times C$.
  For instance, state $\st{3}$ in the figure can transition to state $\st{4}$ or $\st{5}$ or to itself, whereas $\st{5}$ cannot transition anywhere.
  A transition system is specified by a map
  \[
    c\colon C\to \Powf(C)
  \]
  where $\Powf(C)$ is the set of finite subsets of $C$. This maps sends every location $q$ to the set of locations $c(q)\subseteq C$ to which a transition exists.

  \item A Markov chain consists of a set of states, and for each state a probability distribution over all states describes the transition behaviour.
  That is, for each pair of states $q, q'\in C$, the probability $p_{q,q'}\in [0,1]$ denoting the probability to transition from $q$ to $q'$.
  We also attach a boolean label to each state (again, indicated by double border).
  For instance, state $\st{1}$ in the figure steps to state $\st{2}$ with probability $\frac{1}{3}$ and to state $\st{3}$ with probability $\frac{2}{3}$.
  Such a Markov chain is specified by a map
  \[
    c\colon C\to \set{\FALSE,\TRUE}\times \Dist(C)
  \]
  where $\Dist(C)$ is the set of finite probability distributions over $C$.
\end{itemize}

\begin{figure}
  \begin{tabular}{@{}c|c|c|c|@{}}
    & DFA & Transition system & Markov chain \\ \cline{2-4}
    & \begin{tikzpicture}[node distance=1.5cm,baseline=(current bounding box.north)]
      \node[state] (s1) {$\st{1}$};
      \node[state, below left of=s1] (s2) {$\st{2}$};
      \node[state, below right of=s1] (s3) {$\st{3}$};
      \node[state, accepting, below of=s2] (s4) {$\st{4}$};
      \node[state, accepting, below of=s3] (s5) {$\st{5}$};
      \draw
        (s1) edge[arr,above] node{a} (s2)
        (s1) edge[arr,above] node{b} (s3)
        (s2) edge[arr,left] node{a} (s4)
        (s2) edge[arr,above] node{b} (s3)
        (s3) edge[arr,right] node{a} (s5)
        (s3) edge[arr,loop right] node{b} (s3)
        (s4) edge[arr,loop left] node{b} (s4)
        (s4) edge[arr,below,bend right=40] node{a} (s5)
        (s5) edge[arr,below,bend right=15] node{a} (s4)
        (s5) edge[arr,above,bend right] node{b} (s4);
    \end{tikzpicture} &
    \begin{tikzpicture}[node distance=1.5cm,baseline=(current bounding box.north)]
      \node[state] (s1) {$\st{1}$};
      \node[state, below left of=s1] (s2) {$\st{2}$};
      \node[state, below right of=s1] (s3) {$\st{3}$};
      \node[state, below of=s2] (s4) {$\st{4}$};
      \node[state, below of=s3] (s5) {$\st{5}$};
      \draw
        (s1) edge[arr,above] (s2)
        (s1) edge[arr,above] (s3)
        (s1) edge[arr,above] (s4)
        (s2) edge[arr,bend left] (s1)
        (s2) edge[arr,left] (s4)
        (s3) edge[arr,right] (s5)
        (s3) edge[arr,loop right] (s3)
        (s3) edge[arr] (s4)
        (s4) edge[arr,loop left] (s4)
        (s4) edge[arr] (s5);
    \end{tikzpicture} &
    \begin{tikzpicture}[node distance=1.5cm,baseline=(current bounding box.north)]
      \node[state] (s1) {$\st{1}$};
      \node[state, below left of=s1] (s2) {$\st{2}$};
      \node[state, below right of=s1] (s3) {$\st{3}$};
      \node[state, accepting, below of=s2] (s4) {$\st{4}$};
      \node[state, below of=s3] (s5) {$\st{5}$};
      \draw
        (s1) edge[arr,above] node{$\tfrac{1}{3}$} (s2)
        (s1) edge[arr,above] node{$\tfrac{2}{3}$} (s3)
        (s2) edge[arr,left] node{$\tfrac{1}{2}$} (s4)
        (s2) edge[arr,loop left] node{$\tfrac{1}{2}$} (s2)
        (s3) edge[arr,right] node[yshift=-0.15cm]{$\tfrac{1}{2}$} (s4)
        (s3) edge[arr,right,above] node[yshift=-0.05cm]{$\tfrac{1}{4}$} (s2)
        (s3) edge[arr,right,bend right] node[xshift=-0.05cm]{$\tfrac{1}{4}$} (s5)
        (s5) edge[arr,right,bend right] node[xshift=-0.05cm]{$\tfrac{1}{2}$} (s3)
        (s5) edge[arr,below] node[yshift=0.05cm]{$\tfrac{1}{2}$} (s4)
        (s4) edge[arr,loop left] node{$1$} (s4);
    \end{tikzpicture} \\ \hline
  Functor &
    $F(X) = \{\FALSE,\TRUE\} \times X \times X$ &
    $F(X) = \Powf(X)$ &
    $F(X) = \{\FALSE,\TRUE\} \times \Dist(X)$ \\ \hline
  \parbox{2cm}{Coalgebra \\ $c\colon C \to F(C)$} &
  \parbox{2cm}{
    \medskip
    $\st{1} \mapsto (\FALSE,\st{2},\st{3})$ \\
    $\st{2} \mapsto (\FALSE,\st{4},\st{3})$ \\
    $\st{3} \mapsto (\FALSE,\st{5},\st{3})$ \\
    $\st{4} \mapsto (\TRUE,\st{5},\st{4})$ \\
    $\st{5} \mapsto (\TRUE,\st{4},\st{4})$
    \medskip
  } &
  \parbox{2cm}{
    \medskip
    $\st{1} \mapsto \{\st{2},\st{3},\st{4}\}$ \\
    $\st{2} \mapsto \{\st{1},\st{4}\}$ \\
    $\st{3} \mapsto \{\st{3},\st{4},\st{5}\}$ \\
    $\st{4} \mapsto \{\st{4},\st{5}\}$ \\
    $\st{5} \mapsto \{\ \}$
    \medskip
  } &
  \parbox{3.6cm}{
    \medskip
    $\st{1} \mapsto (\FALSE,\{\st{2}\!:\! \tfrac{1}{3}, \st{3}\!:\! \tfrac{2}{3}\})$ \\
    $\st{2} \mapsto (\FALSE,\{\st{2}\!:\! \tfrac{1}{2}, \st{4}\!:\! \tfrac{1}{2}\})$ \\
    $\st{3} \mapsto (\FALSE,\{\st{2}\!:\!\tfrac{1}{4},\st{4}\!:\!\tfrac{1}{2},\st{5}\!:\!\tfrac{1}{4}\})$ \\
    $\st{4} \mapsto (\TRUE,\{\st{4}\!:\! 1\})$ \\
    $\st{5} \mapsto (\FALSE,\{\st{3}\!:\! \tfrac{1}{2}, \st{4}\!:\! \tfrac{1}{2}\})$
    \medskip
  } \\ \hline
  Equivalence &
  $\st{2} \equiv \st{3}, \st{4} \equiv \st{5}$ &
  $\st{1} \equiv \st{2}, \st{3} \equiv \st{4}$ &
  $\st{2} \equiv \st{3} \equiv \st{5}$
  \\ \hline
  \parbox{2cm}{Minimized \\ $c'\colon C' \!\!\to\! F(C')$\!} &
  \parbox{2cm}{
    \medskip
    $\st{1} \mapsto (\FALSE,\st{2},\st{2})$ \\
    $\st{2} \mapsto (\FALSE,\st{4},\st{2})$ \\
    $\st{4} \mapsto (\TRUE,\st{4},\st{4})$
    \medskip
  } &
  \parbox{2cm}{
    \medskip
    $\st{1} \mapsto \{\st{1},\st{3}\}$ \\
    $\st{3} \mapsto \{\st{3},\st{5}\}$ \\
    $\st{5} \mapsto \{\ \}$
    \medskip
  } &
  \parbox{3.6cm}{
    \medskip
    $\st{1} \mapsto (\FALSE,\{\st{2}\!:\! 1\})$ \\
    $\st{2} \mapsto (\FALSE,\{\st{2}\!:\! \tfrac{1}{2}, \st{4}\!:\! \tfrac{1}{2}\})$ \\
    $\st{4} \mapsto (\TRUE,\{\st{4}\!:\! 1\})$
    \medskip
  } \\ \hline
  \end{tabular}
  \caption{Examples of different system types and their encoding as coalgebras for the state set $C = \set{\st{1}, \st{2}, \st{3}, \st{4}, \st{5}}$.}
  \label{fig:coalgexamples}
\end{figure}

We call the data $c(q)$ attached to a state $q$ the \textbf{successor structure} of the state $q$.

\textbf{By generalizing the pattern above, different types of automata can be treated in a uniform way}:
In all these examples, we have a set of states $C$ (where $C = \{\st{1},\st{2},\st{3},\st{4},\st{5}\}$ in the figure),
and then a map $c\colon C\to F(C)$ for the successor structures, for some construction $F$ turning the set of states $C$ into another set $F(C)$.
Such a mapping $F\colon \Set{} \to \Set{}$ (in programming terms one should think of $F$ as a type constructor) is called a functor,
and describes the automaton type.
This point of view allows us to easily consider variations,
such as labelled transition systems, given by $F(X) = \Powf(\{a,b\} \times X)$,
and Markov chains where the states are not labelled but the transitions are labelled,
given by $F(X) = \Dist(\{a,b\} \times X)$.
Other examples, such as monoid weighted systems, Markov Decision processes, and tree automata, are given in \Cref{sec:formalcoalg}.
Representing an automaton of type $F$ by attaching a successor structure of type $F(C)$ to each state $q \in C$ brings us to the following definition:

\begin{definition}
  An automaton of type $F$, or finite $F$-coalgebra, is a pair $(C,c)$ of a finite set of states $C$,
  and a function $c \colon C \to F(C)$ that attaches the successor structure of type $F(C)$ to each state in $C$.
\end{definition}
Since $C$ is a finite set of states, we can give such a map $c$ by listing what each state in $C$ maps to.
For the concrete automata in \Cref{fig:coalgexamples},
the representation using such a mapping $c \colon C \to F(C)$ is given in the \textqt{Coalgebra} row.

\subsection{Behavioural equivalence of states in $F$-automata, generically}

We now know how to uniformly \emph{represent} an automaton of type $F$,
but \textbf{we need a uniform way to state what it means for states to be equivalent}.
Intuitively, we would like to say that two states are equivalent
if the successor structures attached to the two states by the map $c \colon C \to F(C)$ are equivalent.
The difficulty is that the successor structure may itself contain other states,
so equivalence of states requires equivalence of successor structures and vice versa.

A way to cut this knot is to consider a \emph{proposed} equivalence of states,
and then define what it means for this equivalence to be valid, namely:
an equivalence of states is \emph{valid} if proposed to be equivalent states have equivalent successor structures,
where equivalence of the successor structures is considered up to the \emph{proposed} equivalence of states.
In short, the proposed equivalence should be compatible with the transition structure specified by the successor structures.

Rather than representing a proposed equivalence as an equivalence relation $R\subseteq C\times C$ on the state space $C$,
it is better to use a surjective map $r\colon C\to C'$ that assigns to each state a canonical representative in $C'$ identifying its equivalence class (also called \emph{block}).
That is, two states $q,q'$ are equivalent according to $r$, if $r(q) = r(q')$.
Intuitively, $r$ partitions the states into blocks or equivalence classes $\set{q\in C\mid r(q) = y} \subseteq C$ for each canonical representative $y \in C'$.
Not only does this representation of the equivalence avoid quadratic overhead in the implementation,
but it is also more suitable to state the stability condition:

An equivalence $r\colon C \to C'$ is \emph{stable},
if for every two equivalent states $q_1,q_2$ (i.e., with $r(q_1) = r(q_2)$),
the successor structures $c(q_1)$ and $c(q_2)$ attached to the states become \emph{equal}
after replacing states $q$ inside the successor structures with their canonical representative $r(q)$.

This guarantees that we can build a minimized automaton with the canonical representatives $r(q)\in C'$ as state space.
If we do this replacement for both the source and the target of all transitions, we obtain a potentially smaller automaton
$c'\colon C' \to F(C')$.

In order to gain intuition about this, let us investigate our three examples in \Cref{fig:coalgexamples}:
\begin{itemize}
  \item In the finite automaton, the states $\st{4} \equiv \st{5}$ and $\st{2} \equiv \st{3}$ can be shown to be equivalent, so we have $C'=\set{\st{1}, \st{2}, \st{4}}$ and $r\colon C\to C'$ with $\st{3}\mapsto \st{2}$ and $\st{5}\mapsto \st{4}$ (and also $\st{1}\mapsto \st{1}$, $\st{2}\mapsto \st{2}$, $\st{4}\mapsto \st{4}$, which we will use implicitly in future examples).
  We can check that this equivalence is compatible with $c$ by verifying
  that the successor structures of supposedly equivalent states become \emph{equal} after substituting $\st{5} \mapsto \st{4}$ and $\st{3} \mapsto \st{2}$.
  After substituting $\st{5} \mapsto \st{4}$ we indeed have that $c(\st{2}) = (\FALSE,\st{4},\st{3})$ and $c(\st{3}) = (\FALSE,\st{5},\st{3})$ become equal,
  and that $c(\st{4}) = (\TRUE,\st{5},\st{4})$ and $c(\st{5}) = (\TRUE,\st{4},\st{4})$ become equal.
  So this equivalence is stable.
  \item For the transition system, the states $\st{3} \equiv \st{4}$ are equivalent, and $\st{1} \equiv \st{2}$ are equivalent.
  We can verify, for instance, that states $c(\st{1}) = \{\st{2},\st{3},\st{4}\}$ and $f(\st{2}) = \{ \st{1}, \st{4} \}$ are equivalent,
  because after substituting $\st{4} \mapsto \st{3}$ and $\st{2} \mapsto \st{1}$, we indeed have $\{\st{1},\st{3},\st{3}\} = \{ \st{1}, \st{3} \}$,
  because duplicates can be removed from sets.
  Note that it is important that the data for transition systems are sets rather than lists or multisets.
  Multisets also give a valid type of automaton, but they do not give the same notion of equivalence.
  \item For the Markov chain, we can verify $\st{2} \equiv \st{3} \equiv \st{5}$.
  Consider that all three of these states step to state $\st{4}$ with probability $\frac{1}{2}$.
  With the remaining probability $\frac{1}{2}$ these states step to one of the states $\st{2} \equiv \st{3} \equiv \st{5}$, i.e.~they stay in this block.
  State $\st{3}$ steps to either state $\st{2}$ or $\st{5}$ with probability of $\frac{1}{4}$ each. If we however assume that
  state $\st{5}$ behaves equivalent to $\st{2}$, then the branching of state $\st{3}$ is the same as going to state $\st{2}$ with probability
  $\frac{1}{4} + \frac{1}{4} = \frac{1}{2}$ directly.
  Thus, when
  substituting $\st{5} \mapsto \st{2}$ and $\st{3} \mapsto \st{2}$
  the distribution $c(\st{3}) = (\FALSE,\{\st{2}\colon \frac{1}{4},\st{4}\colon\frac{1}{2},\st{5}\colon\frac{1}{4}\})$,
   collapses to $(\FALSE,\{\st{2}\colon\frac{1}{2},\st{4}\colon\frac{1}{2}\})$.
  In other words, edges to equivalent states get merged by summing up their probability.
\end{itemize}

Here we assumed that we were given an equivalence, which we check to be stable.
Our next task is to determine how to find the maximal stable equivalence.
We shall see that this only requires a minor modification to checking that
a given equivalence is stable:
if we discover that an equivalence is \emph{not} stable,
we can use that information to iteratively refine the equivalence until it is stable.

\subsection{Minimizing $F$-automata, generically: the naive algorithm}
\label{sec:minimizing_automata}

In this section we describe a \textbf{naive but generic method for minimizing $F$-automata}\tw{only $F$-coalgebra and \textqt{automata of type $F$} is introduced so far} \cite{KonigKupper14}.
The method is based on the observation that we can start by optimistically assuming that \emph{all} states are equivalent,
and then use the stability check described in the preceding section to determine how to split up into finer blocks.
By iterating this procedure we will arrive at the minimal automaton.

Let us thus see what happens if we blindly assume \emph{all} states to be equivalent,
and perform the substitution where we change every state to state $\st{1}$.
For the finite automaton in \Cref{fig:coalgexamples}, we get
\begin{align*}
  \st{1} &\mapsto (\FALSE,\st{1},\st{1}) &
  \st{2} &\mapsto (\FALSE,\st{1},\st{1}) &
  \st{3} &\mapsto (\FALSE,\st{1},\st{1}) &
  \st{4} &\mapsto (\TRUE,\st{1},\st{1}) &
  \st{5} &\mapsto (\TRUE,\st{1},\st{1})
\end{align*}
Clearly, even though we assumed all states to be equivalent,
the states $\st{1},\st{2},\st{3}$ are still distinct from $\st{4},\st{5}$ because the former three are not accepting whereas the latter two are.
Therefore, even if we initially assumed all states to be equivalent, we discover inequivalent states.
Let us thus try the equivalence $\st{1} \equiv \st{2} \equiv \st{3}$ and $\st{4} \equiv \st{5}$,
and apply substitution where we send $\st{2} \mapsto \st{1}$, $\st{3} \mapsto \st{1}$ and $\st{5} \mapsto \st{4}$:
\begin{align*}
  \st{1} &\mapsto (\FALSE,\st{1},\st{1}) &
  \st{2} &\mapsto (\FALSE,\st{4},\st{1}) &
  \st{3} &\mapsto (\FALSE,\st{4},\st{1}) &
  \st{4} &\mapsto (\TRUE,\st{4},\st{4}) &
  \st{5} &\mapsto (\TRUE,\st{4},\st{4})
\end{align*}
We have now discovered \emph{three} distinct blocks of states: state $\st{1}$, states $\st{2} \equiv \st{3}$ and states $\st{4} \equiv \st{5}$.
If we apply a substitution for \emph{that} equivalence, we get:
\begin{align*}
  \st{1} &\mapsto (\FALSE,\st{2},\st{2}) &
  \st{2} &\mapsto (\FALSE,\st{4},\st{2}) &
  \st{3} &\mapsto (\FALSE,\st{4},\st{2}) &
  \st{4} &\mapsto (\TRUE,\st{4},\st{4}) &
  \st{5} &\mapsto (\TRUE,\st{4},\st{4})
\end{align*}
We did not discover new blocks; we still have three distinct blocks of states:
$\st{1}$, states $\st{2} \equiv \st{3}$ and states $\st{4} \equiv \st{5}$.
Hence, there is no need to change the substitution map sending each state to a representative in the $\equiv$-class,
and so we reached a fixed point.
We can now read off the minimized automaton by deleting states $\st{3}$ and $\st{5}$ from the last automaton above.

\newcommand{\excolwidth}{0.3}
\begin{figure}
  \begin{minipage}{\excolwidth\textwidth}
    \begin{flalign*}
      \st{1} &\mapsto (\FALSE,\st{1},\st{1})& \\
      \st{2} &\mapsto (\FALSE,\st{1},\st{1})& \\
      \st{3} &\mapsto (\FALSE,\st{1},\st{1})& \\
      \st{4} &\mapsto (\TRUE,\st{1},\st{1})& \\
      \st{5} &\mapsto (\TRUE,\st{1},\st{1})&
    \end{flalign*}
  \end{minipage}
  \begin{minipage}{\excolwidth\textwidth}
    \begin{flalign*}
      \st{1} &\mapsto (\FALSE,\st{1},\st{1})& \\
      \st{2} &\mapsto (\FALSE,\st{4},\st{1})& \\
      \st{3} &\mapsto (\FALSE,\st{4},\st{1})& \\
      \st{4} &\mapsto (\TRUE,\st{4},\st{4})& \\
      \st{5} &\mapsto (\TRUE,\st{4},\st{4})&
    \end{flalign*}
  \end{minipage}
  \begin{minipage}{\excolwidth\textwidth}
    \begin{flalign*}
      \st{1} &\mapsto (\FALSE,\st{2},\st{2})& \\
      \st{2} &\mapsto (\FALSE,\st{4},\st{2})& \\
      \st{3} &\mapsto (\FALSE,\st{4},\st{2})& \\
      \st{4} &\mapsto (\TRUE,\st{4},\st{4})& \\
      \st{5} &\mapsto (\TRUE,\st{4},\st{4})&
    \end{flalign*}
  \end{minipage}
  \medskip
  \\ \hrule
  \begin{minipage}{\excolwidth\textwidth}
  \begin{flalign*}
    \st{1} &\mapsto \{\st{1}\}& \\
    \st{2} &\mapsto \{\st{1}\}& \\
    \st{3} &\mapsto \{\st{1}\}& \\
    \st{4} &\mapsto \{\st{1}\}& \\
    \st{5} &\mapsto \{\ \}&
  \end{flalign*}
  \end{minipage}
  \begin{minipage}{\excolwidth\textwidth}
  \begin{flalign*}
    \st{1} &\mapsto \{\st{1}\}& \\
    \st{2} &\mapsto \{\st{1}\}& \\
    \st{3} &\mapsto \{\st{1},\st{5}\}& \\
    \st{4} &\mapsto \{\st{1},\st{5}\}& \\
    \st{5} &\mapsto \{\ \}&
  \end{flalign*}
  \end{minipage}
  \begin{minipage}{\excolwidth\textwidth}
  \begin{flalign*}
    \st{1} &\mapsto \{\st{1},\st{3}\}& \\
    \st{2} &\mapsto \{\st{1},\st{3}\}& \\
    \st{3} &\mapsto \{\st{3},\st{5}\}& \\
    \st{4} &\mapsto \{\st{3},\st{5}\}& \\
    \st{5} &\mapsto \{\ \}&
  \end{flalign*}
  \end{minipage}
  \medskip
  \\ \hrule
  \begin{minipage}{\excolwidth\textwidth}
    \begin{flalign*}
      \st{1} &\mapsto (\FALSE,\{\st{1}\!: 1\})& \\
      \st{2} &\mapsto (\FALSE,\{\st{1}\!: 1\})& \\
      \st{3} &\mapsto (\FALSE,\{\st{1}\!: 1\})& \\
      \st{4} &\mapsto (\TRUE,\{\st{1}\!: 1\})& \\
      \st{5} &\mapsto (\FALSE,\{\st{1}\!: 1\})&
    \end{flalign*}
  \end{minipage}
  \begin{minipage}{\excolwidth\textwidth}
    \begin{flalign*}
      \st{1} &\mapsto (\FALSE,\{\st{1}\!: 1\})& \\
      \st{2} &\mapsto (\FALSE,\{\st{1}\!: \tfrac{1}{2}, \st{4}\!: \tfrac{1}{2}\})& \\
      \st{3} &\mapsto (\FALSE,\{\st{1}\!:\tfrac{1}{2},\st{4}\!:\tfrac{1}{2}\})& \\
      \st{4} &\mapsto (\TRUE,\{\st{4}\!: 1\})& \\
      \st{5} &\mapsto (\FALSE,\{\st{1}\!: \tfrac{1}{2}, \st{4}\!: \tfrac{1}{2}\})&
    \end{flalign*}
  \end{minipage}
  \begin{minipage}{\excolwidth\textwidth}
    \begin{flalign*}
      \st{1} &\mapsto (\FALSE,\{\st{2}\!: 1\})& \\
      \st{2} &\mapsto (\FALSE,\{\st{2}\!: \tfrac{1}{2}, \st{4}\!: \tfrac{1}{2}\})& \\
      \st{3} &\mapsto (\FALSE,\{\st{2}\!:\tfrac{1}{2},\st{4}\!:\tfrac{1}{2}\})& \\
      \st{4} &\mapsto (\TRUE,\{\st{4}\!: 1\})& \\
      \st{5} &\mapsto (\FALSE,\{\st{2}\!: \tfrac{1}{2}, \st{4}\!: \tfrac{1}{2}\})&
    \end{flalign*}
  \end{minipage}
  \caption{Execution of the naive algorithm for the three automata of \Cref{fig:coalgexamples}.}
  \label{fig:naivealgexamples}
\end{figure}

The reader may observe that the process sketched above is quite general,
and can be used to minimize a large class of automata. The sketch translates into the pseudocode in \Cref{algSketchNaive}.

\begin{algorithm}
\begin{algorithmic}
    \upshape
		\Procedure{NaiveAlgorithm}{automaton}
    \Comment{Finds equivalent states of automaton}
    \State Put all states in one block (\ie assume that all states are equivalent)
    \While{number of blocks grows}
      \State Substitute current block numbers in the successor structures
      \State Split up blocks according to the successor structures
    \EndWhile
    \EndProcedure
\end{algorithmic}
\caption{Sketch of the naive partition refinement algorithm}
\label{algSketchNaive}
\end{algorithm}

The execution trace of this naive algorithm for our three example automata of \Cref{fig:coalgexamples} can be found in \Cref{fig:naivealgexamples}.
What the algorithm only needs is the ability to obtain a canonicalized successor structure
after applying a substitution to the successor states.
In general this may involve some amount of computation.
For instance, for transition systems, a purely textual substitution would lead to $\{\st{1},\st{1},\st{1}\}$
assuming all states are conjectured equivalent in the first step,
and the canonical form of this set is $\{\st{1}\}$.
Note that the states $\st{1}-\st{4}$ all have successor structure $\{\st{1}\}$ in the first step of the algorithm,
but they get distinguished from state $\st{5}$,
which has successor structure $\{\ \}$.

We see that in order to talk about equivalence of states,
and in order to perform minimization, we need a notion of substitution and canonicalization.
As it turns out, this corresponds exactly to the standard definition of functor in category theory (for $\Set{}$):
\begin{definition}
  \label{def:functor}
  $F\colon \Set{} \to \Set{}$ is a functor, if given an $p\colon A \to B$ (i.e., a ``substitution''),
  we have a mapping $F[p]\colon F(A) \to F(B)$.
  Furthermore, this operation must satisfy $F[id] = id$ and $F[p \circ g] = F[p] \circ F[g]$.
\end{definition}

We thus require all automata types to be given by functors in the sense of \Cref{def:functor}.
We can then talk about equivalence of states,
and minimize automata by repeatedly applying this operation $F[p]$ as sketched above.
A more formal naive algorithm will be discussed in \Cref{sec:algFinalChain}.

\subsection{The challenge: a generic \emph{and} efficient algorithm}

The problem with the naive algorithm sketched in \Cref{sec:minimizing_automata}
is that it processes all transitions in every iteration of the main loop.
In certain cases, partition refinement (in general) may take $\CT(n)$ iterations to converge, where $n$ is the number of states.
This can happen, for instance, if the automaton has a long chain of transitions,
so in each iteration, only one state is moved to a different block.
\Cref{fig:nsquared} contains three example automata for which the naive algorithm takes $\CT(n)$ iterations
(provided one generalizes the examples to have $n$ nodes).

Since naive algorithm computes new successor structures for all states in each iteration,
the functor operation is applied $\CO(n^2)$ times in total.
Thus, the challenge we set out to solve is the following:

\tikzstyle{arrsum} = [thick,->,>=stealth,dashed]
\begin{figure}
  \begin{minipage}{0.5\textwidth}
  \begin{tikzpicture}[node distance=1.5cm,baseline=(current bounding box.north)]
    \node[state, accepting] (s1) {$\st{1}$};
    \node[state, right of=s1] (s2) {$\st{2}$};
    \node[state, below right of=s2] (s3) {$\st{3}$};
    \node[state, below of=s3] (s4) {$\st{4}$};
    \node[state, below left of=s4] (s5) {$\st{5}$};
    \node[state, left of=s5] (s6) {$\st{6}$};
    \node[state, above left of=s6] (s7) {$\st{7}$};
    \node[state, above of=s7] (s8) {$\st{8}$};
    \draw
      (s1) edge[arr,loop above] node{a} (s1)
      (s2) edge[arr,loop above] node{a} (s2)
      (s3) edge[arr,loop right] node{a} (s3)
      (s4) edge[arr,loop right] node{a} (s4)
      (s5) edge[arr,loop below] node{a} (s5)
      (s6) edge[arr,loop below] node{a} (s6)
      (s7) edge[arr,loop left] node{a} (s7)
      (s8) edge[arr,loop left] node{a} (s8)
      (s1) edge[arr,above] node{b} (s2)
      (s2) edge[arr,above] node{b} (s3)
      (s3) edge[arr,right] node{b} (s4)
      (s4) edge[arr,below] node{b} (s5)
      (s5) edge[arr,below] node{b} (s6)
      (s6) edge[arr,below] node{b} (s7)
      (s7) edge[arr,left] node{b} (s8)
      (s8) edge[arr,above] node{b} (s1);
  \end{tikzpicture}
  \end{minipage}
  \begin{minipage}{0.2\textwidth}
  \begin{tikzpicture}[node distance=1.3cm,baseline=(current bounding box.north)]
    \node[state] (s1) {$\st{1}$};
    \node[state, below of=s1] (s2) {$\st{2}$};
    \node[state, below of=s2] (s3) {$\st{3}$};
    \node[state, below of=s3] (s4) {$\st{4}$};
    \node[state, below of=s4] (s5) {$\st{5}$};
    \draw
      (s1) edge[arr] (s2)
      (s2) edge[arr] (s3)
      (s3) edge[arr] (s4)
      (s4) edge[arr] (s5);
  \end{tikzpicture}
  \end{minipage}
  \begin{minipage}{0.2\textwidth}
    \begin{tikzpicture}[node distance=1.3cm,baseline=(current bounding box.north)]
      \node[state, accepting] (s1) {$\st{1}$};
      \node[state, below of=s1] (s2) {$\st{2}$};
      \node[state, below of=s2] (s3) {$\st{3}$};
      \node[state, below of=s3] (s4) {$\st{4}$};
      \node[state, below of=s4] (s5) {$\st{5}$};
      \node[state, right of=s1] (s1') {$\st{6}$};
      \node[state, below of=s1'] (s2') {$\st{7}$};
      \node[state, below of=s2'] (s3') {$\st{8}$};
      \node[state, below of=s3'] (s4') {$\st{9}$};
      \node[state, accepting, below of=s4'] (s5') {$\st{0}$};
      \draw
        (s1) edge[arr,left] node{$\tfrac{1}{2}$} (s2)
        (s2) edge[arr,left] node{$\tfrac{1}{2}$} (s3)
        (s3) edge[arr,left] node{$\tfrac{1}{2}$} (s4)
        (s4) edge[arr,left] node{$\tfrac{1}{2}$} (s5);
      \draw
        (s2') edge[arr,right] node{$\tfrac{1}{2}$} (s1')
        (s3') edge[arr,right] node{$\tfrac{1}{2}$} (s2')
        (s4') edge[arr,right] node{$\tfrac{1}{2}$} (s3')
        (s5') edge[arr,right] node{$\tfrac{1}{2}$} (s4');
      \draw
        (s1) edge[arr,above] node{$\tfrac{1}{2}$} (s1')
        (s2') edge[arr,above] node{$\tfrac{1}{2}$} (s2)
        (s3) edge[arr,above] node{$\tfrac{1}{2}$} (s3')
        (s4') edge[arr,above] node{$\tfrac{1}{2}$} (s4)
        (s5) edge[arr,above] node{$\tfrac{1}{2}$} (s5');
    \end{tikzpicture}
  \end{minipage}

  \caption{Examples of shapes of automata on which the naive algorithm runs in $\CT(n^2)$.}
  \label{fig:nsquared}
\end{figure}

\medskip

\qquad \textbf{Can we find an asymptotically and practically efficient algorithm for automaton minimization that uses only the successor structure recomputation operation $F[p]$?}

By using only $F[p]$, we do not impose further conditions on the functor $F$ beside $F[p]$ being computable
Since the algorithm does not inspect $F$ any further, the only condition imposed on the functor
is that $F[p]$ is computable for all substitutions $p$ on the state space.

\subsection{Hopcroft's trick: the key to efficient automaton minimization}

A key part of the solution is a principle often called \textqt{Hopcroft's trick} or \textqt{half the size} trick,
which underlies all known asymptotically efficient automata minimization algorithms.
To understand the trick, consider the following game:
\begin{enumerate}
  \item We start with a set of objects, \eg $\{1,2,3,4,5,6,7,8,9\}$.
  \item We chop the set into two parts arbitrarily, \eg $\{1,3,5,7,9\},\{2,4,6,8\}$.
  \item We select one of the sets, and chop it up arbitrarily again, \eg $\{1,3\},\{5,7,9\},\{2,4,6,8\}$.
  \item We continue the game iteratively (possibly until all sets are singletons).
\end{enumerate}
Once the game is complete, we trace back the history of one particular element, say $3$,
and count how many times it was in the smaller part of a split:

\medskip
\quad \textbf{The number of times an element was part of the smaller half of a split is $\CO(\log n)$.}
\medskip

One can prove this bound by considering the evolution of the size of the set containing the element.
Initially, this size is $n$. Each time the element was part of the smaller part of the split, the size of the surrounding
set gets cut in at least half, which can happen at most $\CO(\log n)$ times before we reach a singleton.

This indicates that for efficient algorithms,
we should make sure that the running time of the algorithm is only proportional to the smaller halves of the splits.
In other words, when we split a block, we have to make sure that we do not loop over the larger half of the split.

A slightly more general bound results from considering a game where we can
split each set into an arbitrary number of parts, rather than $2$:

\medskip
\quad \textbf{The number of times an element was part of \underline{a} smaller part of the split is $\CO(\log n)$.}
\medskip

In this case, ``a smaller part of the split'' is to be understood as any part of the split except the largest part.
Thus, if we split $\{1,2,3,4,5,6,7,8,9\}$ into $\{1,3\},\{5,7,9\},\{2,4,6,8\}$, then $\{1,3\},\{5,7,9\}$ are both considered ``smaller parts'',
whereas $\{2,4,6,8\}$ is the larger part.

In terms of algorithm design, our goal shall thus be that when we do a $k$-way split of a block,
we may do operations proportional to all the $k-1$ smaller parts of the split,
but never an operation proportional to the largest part of the split.

\subsection{A sketch of our generic and efficient algorithm}

We design our algorithm based on the naive algorithm and Hopcroft's trick.
The main problem with the naive algorithm is that it recomputes the successor structures of \emph{all} states at each step.
The reader may already have noticed that many of the successor structures in fact stay the same, and are unnecessarily recomputed.
The successor structure of a state only changes if the block number of one of its successors changes.
\emph{The key to a more efficient algorithm is to minimize the number of times a block number changes, so that successor structure recomputation is avoided as much as possible.}

In the naive algorithm, we see that when we split a block of states into smaller blocks,
we have freedom about which numbers to assign to each new sub-block.
We therefore choose to \emph{keep the old number for the largest sub-block}.
Hopcroft's trick will then ensure that a state's number changes at most $\CO(\log n)$ times.

In order to reduce recomputation of successor structures, our algorithm tracks for each block of states (\ie states with the same block number), which of the states are \emph{dirty}, meaning that at least one of their successors' number changed.
The remaining states in the block are \emph{clean}, meaning that the successors did not change.

Importantly, all clean states of a block \emph{have the same successor structure},
because (A) their successors did not change (B) if their successor structure was different in the last iteration, they would have been placed in different blocks.
Therefore, in order to recompute the successor structures of a block,
it suffices to recompute the dirty states and \emph{one} of the clean states,
because we know that all the clean states have the same successor structure.

This sketch translates into the pseudocode of \Cref{algSketchOpt}.

\begin{algorithm}
  \begin{algorithmic}
      \upshape
      \Procedure{PartRefSetFun}{automaton}
      \Comment{Finds equivalent states of automaton}
      \State Put all states in one block (\ie assume that all states are equivalent)
      \State Mark all states dirty
      \While{number of blocks grows}
        \State Pick a block with dirty states
        \State Compute the successor structures of the dirty states and one clean state
        \State Mark all states in the block clean
        \State Split up the block, keeping the old block number for the largest sub-block
        \State Mark all predecessors of changed states dirty
      \EndWhile
      \EndProcedure
  \end{algorithmic}
  \caption{Sketch of the optimized partition refinement algorithm}
  \label{algSketchOpt}
\end{algorithm}

Let us investigate the complexity of this algorithm in terms of the number of successor structure recomputations.
By Hopcroft's trick, a state's number can now change at most $\CO(\log n)$ times,
since we do not change the block number of the largest sub-block.
Whenever we change a state's number, all the predecessors of that state will need to be marked dirty, and be recomputed.
If we take a more global view, we can see that a recomputation may be triggered for every edge in the automaton,
for each time the number of the destination state of the edge changes.
Therefore, if there are $m$ edges, there will be at most $\CO(m \log n)$ successor structure recomputations, \ie at most $\CO(m \log n)$ calls to the functor operation.

In order to make the algorithm asymptotically efficient in terms of the total number of primitive computation steps,
we must make sure to never do any operation that is proportional to the number of clean states in a block.
Importantly, we must be able to split a block into $k$ sub-blocks without iterating over the clean states.
To do this, we have to devise efficient data structures to keep track of the blocks and their dirty states (\Cref{sec:DataStructures}).

We implement our algorithm (\Cref{sec:OurAlgorithm}) with these data structures and efficient methods for computing the functor operation in our tool, \thetool{}.
When using \thetool{}, the user can either encode their automata using a composition of the built-in functors,
or implement their own functor operation and instantiate the algorithm with that.

\subsubsection*{Practical efficiency of the algorithm}
Previous work on algorithms that apply to classes of functors that support more specialized operations in addition to just the functor operation
can give better asymptotic complexity when one considers more fine-grained accounting than just the number of calls to the functor operation \cite{DorschEA17,concurSpecialIssue,coparFM19,WissmannEA2021}.
Perhaps surprisingly, even though our algorithm is very generic
and doesn't have access to these specialized operations,
our algorithm is much faster than the more specialized algorithm in practice (\Cref{sec:Benchmarks}).

However, the limiting factor in practice is not necessarily time but space.
The aforementioned algorithm requires on the order of 16GB of RAM for minimizing automata with 150 thousand states \cite{coparFM19,WissmannEA2021}.
In order to be able to access more memory, distributed algortithms have been developed \cite{BDM22,BlomOrzan05}.
Using a cluster with 265GB of memory, the distributed algorithm was able to minimize an automaton with 1.3 million states and 260 million edges.
By contrast, \thetool{} is able to minimize the same automaton using only 1.7GB of memory.

The reason is that we do not need any large auxiliary data structures;
most of the 1.7GB is used for storing the automaton itself.
Furthermore, because we only need to compute the functor operation for states in the automaton,
we are able to store the automaton in an efficient immutable binary format.

In the rest of the paper we will first give a more formal definition of bisimilarity in coalgebras (\Cref{sec:formalcoalg}),
we describe how we represent our automata, and which basic operations we need (\Cref{sec:Representation}),
we describe the auxiliary data structures required by our algorithm (\Cref{sec:DataStructures}),
we describe our algorithm and provide complexity bounds (\Cref{sec:OurAlgorithm}),
we show a variety of functor instances that our algorithm can minimize (\Cref{sec:Instances}),
we compare the practical performance to earlier work (\Cref{sec:Benchmarks}),
and we conclude the paper (\Cref{sec:Conclusion}).

\section{Coalgebra and Bisimilarity, Formally}
\label{sec:formalcoalg}

In this section we define formally what it means for two states in a coalgebra to be behaviourally equivalent,
and we give examples to show that behavioural equivalence in coalgebras reduces to known notions of bisimilarity for specific functors.

Recall that we model state-based systems as coalgebras for set functors (\Cref{def:functor}):

\begin{definition}
  An \emph{$F$-coalgebra} consists of a carrier set $C$ and a structure map $c\colon C\to FC$.
\end{definition}
Intuitively, the carrier $C$ of a coalgebra $(C,c)$ is the set of states of the
system, and for each state $x\in C$, the map provides $c(x)\in FC$ that is the
structured collection of successor states of $x$. If $F=\Powf$, then $c(x)$ is
simply a finite set of successor states. The functor determines a canonical
notion of behavioural equivalence.
\begin{definition}
    A \emph{homomorphism} between coalgebras $h\colon (C,c)\to (D,d)$ is a map $h\colon
  C\to D$ with $F[h](c(x)) = d(h(x))$ for all $x\in C$. States $x,y$ in a
  coalgebra $(C,c)$ are \emph{behaviourally equivalent} if there is some
  other coalgebra $(D,d)$ and a homomorphism $h\colon (C,c)\to (D,d)$ such that
  $h(x) = h(y)$.
\end{definition}

\begin{example}
  \label{exCoalgebra}
  We consider coalgebras for the following functors (see also \Cref{tabExCoalgebra}):
  \begin{enumerate}
  \item Coalgebras for $\Powf$ are finitely-branching transition systems and
    states $x,y$ are behaviourally equivalent iff they are bisimilar.
  \item An (algebraic) signature is a set $\Sigma$ together with a map $\ar\colon \Sigma\to
    \N$. The elements of $\sigma\in \Sigma$ are called \emph{operation
    symbols} and $\ar(\sigma)$ is the arity. Every signature induces a functor
  defined by
  \[
    \tilde{\Sigma}X = \{ (\sigma,x_1,\ldots,x_{\ar(\sigma)}) \mid \sigma\in\Sigma,
    x_1,\ldots,x_{\ar(\sigma)}\in X \}
  \]
  on sets and for maps $f\colon X\to Y$ defined by
  \[
    \tilde{\Sigma}[f](\sigma,x_1,\ldots,x_{\ar(\sigma)})
    = (\sigma,f(x_1),\ldots,f(x_{\ar(\sigma)})).
  \]
  A state in a $\tilde{\Sigma}$-coalgebra describes a possibly
  infinite $\Sigma$-tree, with nodes labelled by $\sigma \in \Sigma$
  with $\ar(\sigma)$ many children.
  Two states are behaviourally equivalent iff they describe the same
  $\Sigma$-tree.
  \item Deterministic finite automata on alphabet $A$ are coalgebras for
    the signature $\Sigma$ with 2 operation symbols of arity $|A|$.
    States are behaviourally equivalent iff they accept the same language.
  \item For a commutative monoid $(M,+,0)$, the \emph{monoid-valued} functor $M^{(X)}$ \cite[Def.~5.1]{GummS01}
  can be thought of as $M$-valued distributions over $X$:
    \[
      M^{(X)} := \{\mu \colon X\to M\mid \mu(x)\neq 0 \text{ for only finitely many
      }x\in X\}
    \]
    The map $f\colon X\to Y$ is sent by $M^{(-)}$ to
    \[
      M^{(f)}\colon M^{(X)}\to M^{(Y)}
      \qquad
      M^{(f)}(\mu) = \big(y\mapsto \sum_{x\in X, f(x) = y} \mu(x)\big)
    \]
    Coalgebras for $M^{(-)}$ are
    weighted systems whose weights come from $M$.

    A coalgebra $c\colon C\to M^{(C)}$, sends a state $x\in C$ and another state
    $y\in C$ to a weight $m:=c(x)(y)\in M$ which is understood as the weight of
    the transition $x\xrightarrow{m} y$, where $c(x)(y) = 0$ is understood as
    no transition. The coalgebraic behavioural equivalence captures weighted
    bisimilarity~\cite{Klin09}. Concretely, a weighted bisimulation is an
    equivalence relation $R\subseteq C\times C$ such that for all $x\,R\,y$
    and $z\in C$:
    \[
      \sum_{z\,R\,z'} c(x)(z')
      =\sum_{z\,R\,z'} c(y)(z')
    \]

  \item Taking $M = (\Q,+,0)$, we get that $M^{(X)}$ are linear combinations over $X$.
    If we restrict to the subfunctor $\Dist(X) = \{ f \in \Q_{\geq 0}^{(X)} \mid \sum_{x \in X} f(x) = 1\}$ where the weights are nonnegative and sum to 1,
    we get (rational finite support) probability distributions over $X$.\footnote{
      In models of computation where addition of rational numbers isn't linear time,
      one can restrict to fixed-precision rationals $Q_q = \{ \frac{p}{q} \mid p \in \Z \}$ for some fixed $q \in \N_{>0}$ to obtain our time complexity bound.
    }

  \item For two functors $F$ and $G$, we can consider the coalgebra over their composition $F \circ G$.
    Taking $F = \Powf$ and $G = A\times (-)$, coalgebras over $F \circ G$ are labelled transition systems with strong bisimilarity.
    Taking $F = \Powf$ and $G = \Dist$, coalgebras over $F \circ G$ are Markov
    decision processes with probabilistic bisimilarity
    \cite[Def.~6.3]{LarsenS91},
    \cite[Thm.~4.2]{BARTELS200357}.
    For $F=M^{(-)}$ and $G=\Sigma$ for some signature functor, $FG$-coalgebras
    are weighted tree automata and coalgebraic behavioural equivalence is
    backward bisimilarity~\cite{coparFM19,HoegbergEA09}.
  \end{enumerate}
\end{example}

\begin{table}[t]
  \caption{List of functors, their coalgebras, and the accompanying notion
    of behavioural equivalence. The first five is given in
    \Cref{exCoalgebra}, the last introduced later in \Cref{sec:Instances}.}
  \label{tabExCoalgebra}
  \centering
  \begin{tabular}{@{}lll@{}}
    \toprule
    Functor $F(X)$ & Coalgebras $c\colon C\to FC$
    & Coalgebraic behavioural equivalence
    \\
    \midrule
    $\Powf(X)$ & Transition Systems & (Strong) Bisimilarity \\
    $\Powf(A\times X)$ & Labelled Transition Systems & (Strong) Bisimilarity \\
    $M^{(X)}$ & Weighted Systems (for a monoid $M$)& Weighted Bisimilarity \\
    $\Powf (\Dist(X))$ & Markov Decision Processes & Probabilistic Bisimilarity \\
    $M^{(\tilde{\Sigma} X)}$ & Weighted Tree Automata & Backwards Bisimilarity \\
    \midrule
    $\M(X)$ & Monotone Neighbourhood Frames
    & Monotone Bisimilarity
    \\
    \bottomrule
  \end{tabular}
\end{table}

Sometimes, we need to reason about successors and predecessors of a general
$F$-coalgebra:
\begin{definition}
  \label{sucpred}
  Given a coalgebra $c\colon C\to FC$ and a state $x\in C$, we say that $y\in C$
  is a \emph{successor of $x$} if $c(x)$ is not in the image of
  $Fi_y\colon F(C\setminus\{y\}) \to FC$, where $i_y\colon
  C\setminus\set{y}\monoto C$ is the canonical inclusion. Likewise, $x$ is a
  \emph{predecessor} of $y$, and the \emph{outdegree} of $x$ is the number of
  successors of $x$.
\end{definition}

Intuitively, $y$ is a successor of $x$ if $y$ appears somewhere in the term
that defines $c(x) \in F(C)$, like we did in the \textqt{coalgebra} row in
\Cref{fig:coalgexamples}.
We will access the predecessors in the minimization algorithm, and moreover,
the total and maximum number of successors will be used in the run time
complexity analysis.

\section{Coalgebraic Partition Refinement}
\label{sec:CoalgebraicPartitionRefinement}

In this section we will describe how the coalgebraic notions of the preceding section can be used for automata minimization.

\subsection{Representing Abstract Data}
\label{sec:Representation}
When writing an abstract algorithm, it is crucial for the complexity analysis, how the abstract data is actually represented in memory.
We understand finite sets like the carrier of the input coalgebra as finite cardinals $C \cong \set{0,\ldots,|C|-1}\subseteq \N$,
and a map $f\colon C\to D$ for finite $C$ is represented by an array of length $|C|$.

\subsection*{Coalgebra implementation}
\label{sec:coalginterface}
The coalgebra $c \colon C \to FC$ that we wish to minimize is given to the algorithm
as a black-box, because it only needs to interact with the coalgebra via a specific interface.
Whenever the algorithm comes up with a partition $p\colon C\to
C'$, two states $x,y\in C$ need to be moved to different blocks if $F[p](c(x))
\neq F[p](c(y))$. Hence, the algorithm needs to derive $F[p](c(x))$ for states
of interest $x\in C$. Since all partitions are finite, we can assume
$C'\subseteq \N$, and so for simplicity, we consider partitions as maps $p\colon C\to \N$
with the image $\Im(p) = \{0,\ldots,|C'|-1\}$ and so $F[p](c(x))$ is an element of the set $F\N$.

For the case of labelled transition systems, i.e.~$F(X) = \Powf(A\times X)$,
the binary representation of $F[p](c(x))$ is called the \emph{signature of $x\in C$
with respect to $p$}~\cite{BlomOrzan05}.
This straightforwardly generalizes to arbitrary functors $F$~\cite{BDM22,concurSpecialIssue},
so we reuse the terminology \emph{signature} for the binary encoding of the successor structure of $x\in C$ with respect to the blocks the partition $p$ of the previous iteration.

Beside the signatures, the optimized minimization algorithm needs to be able to determine the predecessors of a state, in order to determine which states to mark dirty.
Formally, we require:

\begin{definition}\label{coalgImpl}
  The \emph{implementation} of an $F$-coalgebra $c\colon C\to FC$ is
  the data $(n,\sig,\pred)$ where:
  \begin{enumerate}
  \item $n\in \N$ is a natural number such that $C\cong \set{0,\ldots,n-1}$
  \item $\sig\colon C\times (C\to \N) \to 2^*$ is a function that given a state and a partition, computes the successor structure of the state (represented a binary data), satisfying for all
  partitions $p\colon C\to \N$ (encoded as an array of size $|C|$) that
  \begin{equation}
    \forall x,y\in C\colon \qquad
    \sig(x,p) = \sig(y,p)
    \quad\Leftrightarrow\quad
    F[p](c(x)) = F[p](c(y))
    \label{eqSig}
  \end{equation}
  \item $\pred\colon C\to \Powf C$ is a function such that $\pred(x)$ contains the predecessors of $x$.
  \end{enumerate}
\end{definition}

Passing such a general interface makes the algorithm usable as a library,
because the coalgebra can be represented in an arbitrary fashion in memory, as
long as the above functions can be implemented.

The equivalence involving $\sig$ \eqref{eqSig} specifies that the binary data
of type $2^*$ returned by $\sig$ is some normalized representation of $F[p](c(x))\in F\N$.
For example, in the implementation for $F=\Powf$, an element of $F\N=\Powf \N$
is a set of natural numbers. Since e.g.~$\set{2,0}$
and $\set{0,2,2} \in \Powf \N$ are the same set, the $\sig$ function
essentially needs to sort the arising sets and remove duplicates:
\begin{example}
  We can represent $\Powf$-coalgebras $c\colon C\to \Powf C$
  by keeping for every state $x\in C$ an array of its successors $c(x) \subseteq C$ in memory.
  As a pre-processing step, we directly compute the predecessors for each state
  $x\in C$ and keep them as an array $\pred(x)\subseteq C$ for every state $x$ in memory as well
  (computing the predecessors of all states can be done in linear time, and thus does not affect the complexity of the algorithm).
  With $n := |C|$, the remaining function $\sig$ is implemented as follows:
  \begin{enumerate}
  \item Given $p\colon C\to \N$ and $x\in C$, create a new array $t$ of
  integers of size $|c(x)|$. For each successor $y\in c(x)$, add $p(y)\in \N$
  to $t$; this runs linearly in the length of $t$ because we assume that the
  map $p$ is represented as an array with $\CO(1)$ access.

  \item Sort $t$ via radix sort and then remove all duplicates, with both
    steps taking linear time.
  \item Return the binary data blob of the integer array $t$.
  \end{enumerate}
  For $\Powf$, the computation of the signature of a state $x\in C$ thus takes
  $\CO(|c(x)|)$ time.
\end{example}
We discuss further instances in \Cref{sec:Instances} later.

\subsection*{Renumber}
By encoding everything as binary data in a normalized way, we are able to make
heavy use of radix sort, and thus achieve linear bounds on sorting tasks.
This trick is also used in the complexity analysis of Kanellakis and Smolka,
who refer to it as lexicographic sorting method by Aho, Hopcroft, and Ullman~\cite{AHU74}.
We use this trick in order to turn arrays of binary data $p\colon B\to 2^*$ into
their corresponding partitions $p'\colon B \to \{0,\ldots,|\Im(p)|-1\}$ satisfying
  $p(x) = p(y)
  \Longleftrightarrow\
  p'(x) = p'(y)
  \text{ for all }x,y\in B$.
The pseudocode is listed in \Cref{algoRenumber}: first, a permutation
$r\colon B\to B$ is computed such that $p\circ r\colon B\to
2^*$ is sorted. This radix sort runs in $\CO(\sizeof{p})$, where $\sizeof{p} = \sum_{x\in B} |p(x)|$ is the total size of the entire array $p$.
Since identical entries in $p$ are now adjacent, a simple for-loop iterates over $r$
and readily assigns block numbers.

\begin{algorithm}[t]
  \caption{Renumbering an array using radix sort}
	\begin{algorithmic}
    \upshape
		\Procedure{Renumber}{$p\colon B\to 2^*$}
    \State Create a new array $r$ of size $|B|$ containing numbers $\range{0}{|B|}$
    \State Sort $r$ by the key $p\colon B\to 2^*$ using radix-sort
    \State Create a new array $p'\colon B\to \N$
    \State $j\gets 0$
    \For{$i \in \range{0}{|B|}$}
    \If{$i > 0$ and $p[r[i-1]] \neq p[r[i]]$} $j\gets j + 1$ \EndIf
    \State $p'[r[i]] \gets j$
    \EndFor
    \State \Return $p'$
    \EndProcedure
  \end{algorithmic}
  \label{algoRenumber}
\end{algorithm}
\todo{Later on we want to assume that renumber(p)[0] = 0.
This is to ensure that the states with the same signature as the clean states are put next to the clean states.
We have to modify the algorithm to ensure that.}

\begin{lemma}
  \label{renumberCorrect}
  \Cref{algoRenumber} runs in time $\CO(\sizeof{p})$ for the parameter $p\colon
  B\to 2^*$ and returns a map $p'\colon B\epito b$ for some $b\in \N$
  such that for all $x,y\in B$ we have
  \(
  p(x) = p(y)
  \Leftrightarrow
  p'(x) = p'(y)
  \).
\end{lemma}
\begin{proofappendix}{renumberCorrect}
  \textbf{Run Time Complexity.}
  Note that the radix sort needs to take care that the bit-strings do not have
  uniform length. This can be easily achieved in linear time (especially because
  the sorting is not required to be stable).

  \textbf{Correctness.} After the sorting operation, the blocks' identical
  elements are adjacent in the permutation $r$. Thus, the final for-loop can
  create a new block whenever it sees an element different from the previous element.
\end{proofappendix}

  In the actual implementation, we use hash maps to implement $\renumber$.
  This is faster in practice but due to the resolving of
  hash-collisions, the theoretical worst-case complexity of the implementation has an additional log
  factor.

The renumbering can be understood as the compression of a map $p\colon B\to 2^*$
to an integer array $p'\colon B\to \N$. In the algorithm, the array elements of
type $2^*$ are encoded signatures of states.

\subsection{The Naive Method Coalgebraically}
\label{sec:algFinalChain}
To illustrate the use of the encoding and notions defined above, let us restate the naive method (\Cref{algSketchNaive}, \cite{KonigKupper14,KanellakisSmolka83}) in \Cref{algFinalChain}.
Recall that the basic idea is that it computes a sequence of partitions
$p_i\colon C\to P_i$ ($i\in \N$) for a given input coalgebra $c\colon C\to FC$.
Initially this partition identifies all states $p_0\colon C\to 1$.
In the first iteration, the map $p'\colon (C\xrightarrow{c} FC\xrightarrow{F[p]}
F\N)$ sends each state to its \emph{output behaviour} (this
distinguishes final from non-final states in DFAs and deadlock from live states
in transition systems).
Then this partition is refined successively under consideration of the transition structure:
$x,y$ are identified by $p_{i+1}\colon C\to P_{i+1}$ iff they are identified by
the composed map
\[
  C\xrightarrow{c} FC
  \xrightarrow{F[p_i]} FP_i.
\]
The algorithm terminates as soon as $p_i = p_{i+1}$, which then identifies
precisely the behaviourally equivalent states in the input coalgebra $(C,c)$.

\begin{algorithm}
  \caption{The naive algorithm, also called \emph{final chain partitioning}}
  \label{algFinalChain}
	\begin{algorithmic}
    \upshape
		\Procedure{NaiveAlgorithm'}{$c\colon C\to FC$}
    \State Create a new array $p\colon C\to \N := (x \mapsto 0)$ \Comment{i.e.~$p[x] = 0$ for all $x\in C$}
    \While{$|\Im(p)|$ changes}
    \State compute $p'\colon C\to 2^* := x \mapsto \sig(x,p)$
    \Comment{$p'[x] \in 2^*$ is the encoding of $F[p](c(x))\in F\N$}
    \State $p\colon C \to \N \gets \renumber(p')$
    \EndWhile
    \EndProcedure
  \end{algorithmic}
\end{algorithm}

Recently, Birkmann \etal~\cite{BDM22} have adapted this algorithm to a
distributed setting, with a run time in $\CO(m\cdot n)$.

\subsection{The Refinable Partition Data Structure}
\label{sec:DataStructures}

For the naive method it sufficed to represent the quotient on the state space
$p\colon C\to \N$ by a simple array. For more efficient algorithms like our \Cref{algSketchOpt}, it is crucial to quickly perform certain operations on the partition, for which
we have built upon a refinable partition data structure~\cite{Valmari09,ValmariLehtinen08}.
The data structure keeps track of the partition of the states into blocks.
A key requirement for our algorithm is the ability to split a block into $k$ sub-blocks, where $k$ is arbitrary.
The refinable partition also tracks for each state whether it is \emph{clean} or \emph{dirty},
and a \emph{worklist} of blocks with at least one dirty state.

Let us define the exposed functionality of the refinable partition data structure:
\begin{enumerate}
  \item Given (the natural number identifying) a block $B$, return its dirty states $\BD$ in $\CO(|\BD|)$.
  \item Given a block $B$, return one arbitrary clean state in $\CO(1)$ if there is any. We denote this by the set $\BCone$ of cardinality at most 1. $\BCone$ contains a clean state of $B$
    or is empty if all states of $B$ are dirty.
  \item Return an arbitrary block with a dirty state and remove it from the worklist, in $\CO(1)$.
  \item $\MarkDirty(s)$: mark state $s$ dirty, and put its block on the
    worklist, in $\CO(1)$.
  \item $\Split(B,A)$: split a block $B$ into many sub-blocks according to an array
    $A\colon \BD\to \N$.
    The array $A$ indicates that the $i$-th dirty state is placed in the
    sub-block $A[i]$, meaning that two states $s_1,s_2$ stay together iff $A[s_1] = A[s_2]$.
    The clean states are placed in the $0$-th sub-block, with those states satisfying $A[s] = 0$.

      The block identifier of $B$ gets re-used as the identifier for largest sub-block,
      and all states of $B$ are marked clean. $\Split$ returns the list of all
      newly allocated sub-blocks, i.e.~those except the re-used one.

      For the time complexity of our algorithm, it is important that $\Split(B,A)$
      runs in time $\CO(|\BD|)$, regardless of the number of clean states.
\end{enumerate}
In order to implement these operations with the desired run time complexity, we maintain the following data structures:
\begin{itemize}
  \item $\states$ is an array of size $|C|$ containing all states of $C$. Every
  block is a section of this array, and the
  other stuctures are used to quickly find and update the entries in the $\states$ array. A
  visualization of an extract of this array is shown in \Cref{algMarkDirtySplit}; for example
  lowermost row shows three blocks of size 5, 3, and 1, respectively.
  \item The array $\loc$ is inverse to $\states$; $\loc[s]$ provides the
  index (\textqt{location}) of state $s$ in $\states$.
  \item $\blocks$ is an array of tuples $(start,mid,end)$ and specifies the blocks of the partition.
    A block identifier $B$ is simply an index in this array and $\blocks[B]=(start,mid,end)$ means that
    block $B$ starts at $\states[start]$ and ends before $\states[end]$, as indicated in the visualization in \Cref{algMarkDirtySplit}. The range $\range{start}{mid}$ contains the clean states
    of $B$ and $\range{mid}{end}$ the dirty states. E.g.~$mid=end$ iff the block has no dirty states.
  \item The array $\block$ of size $|C|$ that maps every state $s\in C$ to
    the ID $B = \block[s]$ of its surrounding block.
  \item $\worklist$ is a list of block identifiers and mentions those blocks with at least one dirty state.
\end{itemize}

\newcommand{\pluseq}{\mathrel{+}=}%
\newcommand{\minuseq}{\mathrel{-}=}%
\begin{algorithm}[h]
  \caption{Refinable partition data structure with $n$-way split}
  \label{algMarkDirtySplit}

\begin{minipage}[t]{0.48\textwidth}

  \begin{algorithmic}
    \upshape
    \Procedure{MarkDirty}{$s$}
    \LComment{Determine the block data}
      \State $B := \block[s]$
      \State $j := \loc[s]$
      \State $(start,mid,end) := \blocks[B]$
    \LComment{Do nothing if already dirty}
    \If{$mid \leq j$} \Return \EndIf
    \LComment{Add to worklist if first dirty state}
    \If{$mid = end$} $\worklist.add(B)$ \EndIf
    \LComment{Swap $s$ with the last clean state}
    \State $s' := \states[mid - 1]$
    \State $\loc[s'] := j$
    \State $\loc[s] := mid$
    \State $\states[j] := s'$
    \State $\states[mid] := s$
    \LComment{Move marker to make $s$ dirty}
    \State $\blocks[B].mid \minuseq 1$
    \EndProcedure
  \end{algorithmic}

  \vspace{0mm}
  \centering\hspace*{5mm}
  \def\partitionoverhang{4mm}
  \newcommand{\drawPartitionBlock}[3]{%
      \foreach \statename [count=\x,remember=\statename as \laststate] in {#2} {
        \ifthenelse{1=\x}{
          \node[arrayelem] (#1\statename) {$s_\statename$};
          \global\edef\myfirstname{#1\statename}
        }{
          \node[anchor=west,arrayelem,alias=#1\statename] (#1\statename) at (#1\laststate.east) {$s_\statename$};
          \draw[inner block line] (#1\statename.north west)
          -- (#1\statename.south west);
          \global\edef\mylastname{#1\statename}
        }
      }
      \draw[outer block line] (\myfirstname.north west) -- (\myfirstname.south west);
      \draw[outer block line] (\mylastname.north east) -- (\mylastname.south east);
      \foreach \name [count=\curcount] in {#3} {
        \draw[block separator line] (\name.north east) -- (\name.south east);
      }
      \node[overlay,anchor=east] at ([xshift=-0pt]\myfirstname.west) {$\cdots$};
      \node[overlay,anchor=west] at ([xshift= 0pt]\mylastname.east) {$\cdots$};
      \draw[outer block line] ([xshift=-\partitionoverhang]\myfirstname.north west)
        -- ([xshift=\partitionoverhang]\mylastname.north east);
      \draw[outer block line] ([xshift=-\partitionoverhang]\myfirstname.south west)
      -- ([xshift=\partitionoverhang]\mylastname.south east);
  }
  \begin{tikzpicture}[
    x=14pt,
    arrayelem/.style={
      text depth=0pt,
      inner ysep=5pt,
    },
    start mid end/.style={
      label node/.style={},
      label arrow/.style={
        ->,shorten >= 2pt,
      },
    },
    state range brace/.style={
      decorate,
      decoration={brace,raise=2pt,amplitude=7pt},
      line cap=round,
      draw=docStyleGray,
      line width=1pt,
      every node/.append style={
        yshift=8pt,
        anchor=south,
      },
    },
    inner block line/.style={
      draw=docStyleGray,
      dashed,
    },
    outer block line/.style={
      draw=docStyleBlue,
      line width=1.5pt,
    },
    block separator line/.style={
      outer block line,
      dashed,
    },
    ]
    \begin{scope}[yshift=0cm]
      \drawPartitionBlock{s}{1,...,9}{s4}
      \draw[state range brace] (s1.north west) --
      node {clean} ([xshift=-1pt]s4.north east);
      \draw[state range brace] ([xshift=1pt]s5.north west)
      -- node {dirty states $\BD$} (s9.north east);
      \draw[state range brace] ([yshift=7mm]s1.north west)
      -- node {block $B$} ([yshift=7mm]s9.north east);
      \begin{scope}[start mid end]
        \foreach \labeltext/\anchor/\name in {start/west/s1,end/east/s9,mid/west/s5} {
          \node[label node] (labeltext) at ([yshift=-6mm]\name.south \anchor) {\labeltext};
          \path[label arrow] (labeltext.north) edge (\name.south \anchor);
        }
      \end{scope}
    \end{scope}
    \begin{scope}[yshift=-19mm,
      start mid end/.style={
        label node/.style={
          opacity=0,
          overlay,
        },
        label arrow/.style={
          opacity=0,
          overlay,
        },
      },
      ]
      \drawPartitionBlock{t}{1,2,4,3,5,6,7,8,9}{t4}
      \begin{scope}[yshift=-17mm,
          block separator line/.style={outer block line},
        ]
        \drawPartitionBlock{u}{1,2,4,7,8,3,6,9,5}{u8,u9}
      \end{scope}
      \node[overlay] at ([xshift=0mm,yshift=3mm]t1.north west)
      {\textsc{MarkDirty}($s_3$)};
      \node[overlay] at ([xshift=5mm,yshift=6mm]u1.north west)
      {\textsc{Split}($B$,$[1,2,1,0,0,1]$)};
    \end{scope}
    \begin{scope}
      \foreach \source/\target in {s3/t3,s4/t4,%
        t3/u3,
        t5/u5,
        t6/u6,
        t7/u7,
        t8/u8,
        t9/u9%
      } {
        \draw[->,shorten <= 0mm, shorten >= 0mm,
        line width=1pt,
        line cap=round,
        draw=docStyleGray,]
           ([yshift=-1mm]\source.south) -- ([yshift=1mm]\target.north);
      }
    \end{scope}
  \end{tikzpicture}
\end{minipage}\hfill%
\begin{minipage}[t]{0.52\textwidth}

\newcommand{\countA}{D}
\begin{algorithmic}
  \upshape
  \Procedure{Split}{$B$, $A\colon \BD \to \N$}
  \LComment{Cumulative counts of sub-block sizes}
  \State $(start,mid,end) := \blocks[B]$
    \State $\countA[\range{0}{\max_i A[i]+1}] := 0$
    \State $\countA[0] := mid - start$
    \For{$j \in \BD$}
      \State $\countA[A[j]] \pluseq 1$
    \EndFor
    \State $\imax = \argmax_i \countA[i]$
    \For{$i \in \range{1}{|\countA|}$}
      \State $\countA[i] \pluseq \countA[i-1]$
    \EndFor
  \LComment{Re-order the states by $A$-value}
    \State $\dirtyStates := \operation{copy}(\states[\range{mid}{end}])$
    \For{$i \in \operation{reverse}(\range{0}{|A|})$}
      \State $\countA[A[i]] \minuseq 1$
      \State $j := start + \countA[A[i]]$
      \State $\states[j] := \dirtyStates[i]$
      \State $\loc[\states[i]] := j$
    \EndFor
    \State $\countA[0] \minuseq mid - start$
  \LComment{Create blocks and assign IDs}
    \State $\countA.add(end - start)$
    \State $old\_block\_count := |\blocks|$
    \For{$i \in \range{0}{|\countA|-1}$}
        \State $j_0 := start + \countA[i]$
        \State $j_1 := start + \countA[i+1]$
        \If{$i = \imax$}
          \State $\blocks[B] = (j_0,j_1,j_1)$
        \Else
          \State $\blocks.add(j_0,j_1,j_1)$
          \State $idx := |\blocks|-1$
          \State $\block[\states[\range{j_0}{j_1}]] := idx$
        \EndIf
    \EndFor
  \State \Return $\range{old\_block\_count}{|\blocks|}$
  \EndProcedure
\end{algorithmic}

\end{minipage}
\end{algorithm}

With this data, we can implement the above-mentioned interface:
\begin{enumerate}
\item For a block $B$, its dirty states $\BD$ are the states $\states[\range{mid}{end}]$ where $\blocks[B] = (start,mid,end)$.
\item One arbitrary clean state $\BCone$ of a given block $B$ is determined in a similar fashion: for
  $\blocks[B] = (start,mid,end)$, if $start=mid$, then there is one clean state
  $\BCone = \set{}$, and otherwise we chose $\BCone = \set{\states[start]}$.
\item Returning an arbitrary block containing a dirty state is just a matter of extracting one element from $\worklist$.
\item The pseudocode of $\MarkDirty$ is listed in \Cref{algMarkDirtySplit}:
  when marking a state $s\in C$ dirty, we first find the boundaries
  $(start,mid,end) = \blocks[B]$ of the surrounding block $B = \block[s]$. By the index $\loc[s]$, we can check in $\CO(1)$ whether $s$ is in the first (\textqt{clean}) or second (\textqt{dirty}) part of the block. Only if $s$ wasn't dirty already, we need to do something: if $B$ did not contain dirty states yet ($start=mid$), $B$ now needs to be added to the $\worklist$. Then, we change the location of $s$  in the main array such that it becomes the last clean state, and then we make it dirty by moving the decrementing the index $mid$.

  In the example in \Cref{algMarkDirtySplit}, the content of $\states$ is visualized. The bold dashed line visualizes the $mid$ position, so states on the left of it are clean, states on the right are dirty. The call to $\MarkDirty(s_3)$ transforms the first row into the second row:
  it does so by moving
  $s_3$ from the clean states of $B$ to the dirty ones, while $s_4$ stays
  clean.
\item The pseudocode of $\Split$ is listed in \Cref{algMarkDirtySplit}: for a
  block $B$, the caller provides us with an array $A\colon \BD\to \N$ that
  specifies which of the states stay together and which are moved to separate
  blocks. In the visualized example, $A=[1,2,1,0,0,1]$ represents the map
  \[
      s_3\mapsto 1,\quad
      s_5\mapsto 2,\quad
      s_6\mapsto 1,\quad
      s_7\mapsto 0,\quad
      s_8\mapsto 0,\quad
      s_9\mapsto 1
  \]
  So $\Split(B,A)$ needs to create new blocks $s_3,s_6,s_9$ and $s_5$, while
  $s_7,s_8$ stay with the clean states. In any case, the clean states stay in the same block, so we can understand $A$ as an efficient representation of the map
  \[
      \bar A\colon B\to \N \qquad
      \bar A(s) = \begin{cases}
        A(s) &\text{if }s\in \BD,\\
        0 &\text{otherwise.}
      \end{cases}
  \]
  Then, two states $s,s'\in B$ stay in the same block iff $\bar A[s] = \bar A[s']$.
  In the implementation, we first create an auxiliary array $D$ which has
  different meanings. Before the definition of $\imax$, it counts the sizes of
  the resulting blocks:
  \[
      D[i] = \set{j\in B \mid \bar A[j] = i}.
  \]
  We compute $D$ by initializing $D[0]$ with the number of clean states
  ($mid-start$) and iterating over $A$. The index of the largest block
  remembered in $\imax$, and then we change the meaning of $D$ such that it
  now holds partial sums $D[i] := \sum_{0\le j < i} D[j]$.
  For every new block $i$, this sum $D[i]$ denotes the end of the block,
  relative to the start of the old block $B$.

  We use the sums to re-order the states such that states belonging to the same sub
  block come next to each other. The for-loop moves every state $i\in B$ to the
  end of the new block $A[i]$ and decrements $D[A[i]]$ such that the next state
  belonging to $A[i]$ is inserted before that.
  Finally, we do not need to move the clean states to sub-block $0$, so we
  simply decrement $D[0]$ by the number of clean states. Since we have inserted
  all the elements at the end of their future subblocks and have decremented
  the entry of $D$ during each insertion, the entries of $D$ now point to the
  \emph{first element} of each future subblock.

  Having the states in the right position within $B$, we can now create the
  subblocks with the right boundaries. For convenience, we add the (relative)
  end of $B$ to $D$, because then, every sub block $i$ ranges from $D[i]$ to
  $D[i+1]$. We had saved the index of the largest subblock $\imax$, which will
  inherit the block identifier of $B$ and the entry $\blocks[B]$. For all other
  subblocks, we add a new block to $\blocks$. All new blocks have no dirty
  states, so $mid=end$ for the new entries. If we have added a new block, then
  we need to update $\block[s]$ for every state $s$ in the subblock.
\end{enumerate}

\subsection{Optimized Algorithm}
\label{sec:OurAlgorithm}

With the refinable partition data structure at hand, we can improve on the naive algorithm without restricting the choice of $F$.
Our efficient algorithm is given in \Cref{algPartRefSetFun}.
We start by creating a refinable partition data structure with a single block for all the states.
We then iterate while there is still a block with dirty states, i.e.~with states
whose signatures should be recomputed.
We split the block into sub-blocks in a refinement step that is similar to the naive algorithm, and re-use the old block for the largest sub-block.

To achieve our complexity bound, this splitting must happen in time $|\BD|$,
regardless of the number of clean states.
Fortunately, this is possible because the clean states all have the same
signature, because all their successors remained unchanged. Hence, it suffices to
compute the signature for one arbitrary clean state, denoted by $\BCone$.
Depending on the functor, it might happen that there are dirty states $d\in
\BD$ that have the same signature as the clean states. Having marked a state as
\textqt{dirty} just means that the signature might have changed compared to the
previous run, so it might be that the signature of a dirty state turns out to
be identical to the clean states in the block $B$.

The wrapper $\renumber'$ then first compresses $p\colon \BD\to 2^*$ to
$A\colon \BD\to \N$. Then, $\renumber'$ ensures that those dirty states $d\in\BD$ with
the same signature as the clean states satisfy $A(d) = 0$. This is used in
$\Split$: in the splitting operation, two dirty states $d,d' \in \BD$ stay in the same block iff $A(d) = A(d')$ and the clean states end up in the same block as the dirty states $d$ with $A(d) = 0$.

After the block $B$ is split, we need to mark all states
$x\in B$ as dirty whose signature might have possibly changed due to the
updated partition. If the successor $y$ of $x\in B$ was moved to a new block,
i.e.~if $p(y)$ changed, this might affect the signature of $x$. Conversely,
if no successor of $x$ changed block, then the signature of $x$ remains unchanged:
\begin{lemma}
  \label{sameSigForCleans}
  If for a finite coalgebra $c\colon C\to FC$, two partitions
  $p_1,p_2\colon C\to \N$
  satisfy $p_1(y) = p_2(y)$ for all successors $y$ of $x\in C$,
  then $F[p_1](c(x)) = F[p_2](c(x))$.
\end{lemma}
\begin{proofappendix}{sameSigForCleans}
  Let $S\subseteq C$ be the subset of all successors of $x$.
  Since $C$ is finite, we have the finite intersection
  \[
    S = \bigcap \set[\big]{C\setminus\{y\}\mid y\text{ not a successor of }x}.
  \]
  In general, every set functor preserves finite non-empty intersections~\cite{trnkova69}. We distinguish cases:
  \begin{enumerate}
  \item Case $S \neq \emptyset$: If $S$ is non-empty, then the above intersection is preserved by $F$, so
  \[
    FS = \bigcap \set[\big]{F(C\setminus\{y\})\mid y\text{ not a successor of }x}.
  \]
  We have that $c(x)$ is in the image of every $F(C\setminus\{y\})\monoto FC$ for every
  non-successor $y$ of $x$. Since $F$ preserves the above intersection, we have
  that $c(x)$ is also in the image of $Fs\colon FS\monoto FC$ (where $s\colon S\monoto C$ is just
  the inclusion map).

  Since $p_1(z) = p_2(z)$ for all $z\in S$ by assumption, the domain restrictions of $p_1,p_2\colon C\to \N$ to $S\subseteq C$ are identical:
  $p_1|_S =  p_2|_S\colon S\to \N$. Then, we can conclude $F[p_1](c(x)) = F[p_2](c(x))$ by the diagram:
  \[
    \begin{tikzcd}
      1
      \arrow{r}{c(x)}
      \arrow{dr}[swap]{\exists c'}
      & FC \arrow[shift left=1]{r}{F[p_1]}
      \arrow[shift right=1]{r}[swap]{F[p_2]}
      &[13mm] F\N
      \\
      & FS
      \arrow[>->]{u}[swap]{Fs}
      \arrow[bend right=10]{ur}[sloped,below]{F(p_1|_S) = F(p_2|_S)}
    \end{tikzcd}
  \]

  \item Case $S=\emptyset$ and $|C| = 1$. This implies that $C = \set{x}$ and, because $S=\emptyset$ entails that $x$ is not a successor of itself. Hence, there is some map $c'\colon 1\to F(C\setminus\set{x})$ making
  \[
    \begin{tikzcd}[ampersand replacement=\&]
      1
      \arrow{r}{c(x)}
      \arrow{dr}[swap]{\exists c'}
      \& FC
      \\
      \& F(C\setminus\set{x}) \rlap{\ensuremath{~= F\emptyset = FS}}
      \arrow[>->]{u}[swap]{Fs}
    \end{tikzcd}
  \]
  commute. The rest of the reasoning is identical to the first case.

  \item Case $S=\emptyset$ and $|C|\neq 1$. Since $x\in C$, we have $|C|\ge 2$ and so $C\setminus\set{y}$ is not empty for all $y\in C$.
  We now switch from $F\colon \Set\to\Set$ to its Trnkov\'a-hull $\bar F\colon \Set\to\Set$~\cite{trnkova71}. The functor $\bar F$ coincides with $F$ on all non-empty sets
  \[
    X\neq \emptyset \quad\Rightarrow\quad
    \bar FX = FX
  \]
  and on maps with non-empty domain, and has the property that it preserves all finite intersections (also the possibly empty ones).
  In particular, we have $\bar F\N = F\N$, $\bar FC = FC$,
  and $\bar F(C\setminus\set{y}) = F(C\setminus\set{y})$
  for all $y\in C$.
  So whenever $y$ is not a successor of $x$ in the original coalgebra $c\colon C\to FC$,  we have that $c(x)\colon 1\to FC=\bar FC$ factors through the canonical injection
  \[
    \bar F(C\setminus\set{y}) \monoto \bar FC.
  \]
  Since $\bar F$ preserves the (empty) intersection, we have
  \[
      \bar F S = \bigcap \set{\bar F(C\setminus\set{y}\mid \text{$y$ is not a successor of $x$}}
  \]
  and there is some $c'$ with
  \[
    \begin{tikzcd}[ampersand replacement=\&]
      1
      \arrow{r}{c(x)}
      \arrow{dr}[swap]{\exists c'}
      \& \bar FC\rlap{\ensuremath{~=FC}}
      \\
      \& \bar FS.
      \arrow[>->]{u}
    \end{tikzcd}
  \]
  The rest of the reasoning is like in the first case, only with $F$ replaced with $\bar F$.
  \qedhere
  \end{enumerate}
\end{proofappendix}

\begin{algorithm}[t]
  \caption{Optimized Partition Refinement for all Set functors}
  \label{algPartRefSetFun}
	\begin{algorithmic}
    \upshape
		\Procedure{PartRefSetFun}{$C$, $\sig$, $\pred$}
    \Comment{i.e.~for the implementation of $c\colon C\to FC$}
    \State Create a new refinable partition structure $p\colon C\to \N$
    \State Init $p$ to have one block of all states, and all states marked dirty.
    \AlgoSeparator
    \While{there is a block $B$ with a dirty state}
    \BeginBox[fixed width,algo block brace alt={Compute signatures, \\in total $\CO(m \log n)$ calls\\ to the coalgebra}]
      \State Compute the arrays
      \State \quad $(sigs_{\dirty}\colon \BD \to 2^*) := (x\mapsto \sig(x, p))$
      \State \quad $(sigs_{\clean}\colon \BCone \to 2^*) := (x\mapsto \sig(x, p))$
    \EndBox
    \AlgoSeparator
    \BeginBox[fixed width,algo block brace alt={Split $B$ according to \\ signatures in $\CO(|\BD|)$}]
        \State $A \colon \BD\to \N := \renumber'(sigs_{\dirty}, sigs_{\clean})$
      \State $\vec{B}_{new} := \Split(B,A)$
    \EndBox
    \AlgoSeparator
    \BeginBox[fixed width,algo block brace alt={Mark dirty all states with \\ a successor in a new block \\ in total time $\CO(m \log n)$}]
      \For{every $B' \in \vec{B}_{new}$ and $s \in B'$}
        \For{every $s'\in \pred(s)$}
          \State $\MarkDirty(s')$
        \EndFor
      \EndFor
    \EndBox
    \EndWhile
    \AlgoSeparator
    \State \Return the partition $p$
    \EndProcedure
  \end{algorithmic}
  \begin{algorithmic}
    \upshape
		\Procedure{Renumber'}{$p\colon \BD \to 2^{*}, q : \BCone \to 2^{*}$}
    \State $(A\colon \BD\to \N) := \renumber(p)$
    \BeginBox[fixed width,algo block brace alt={Ensure that $A(d) = 0$ for all dirty \\ states $d$ that have the same \\ signature as the clean states.}]
    \If{there are $d\in \BD, c\in \BCone$ with $p(d) = q(c)$}
    \State Swap the values $0$ and $A(d)$ in the array $A$.
    \EndIf
    \EndBox
    \State \Return $A$
    \EndProcedure
  \end{algorithmic}
\end{algorithm}

We can now prove correctness of the partition refinement for coalgebras:
\begin{theorem}
  \label{mainCorrectness}
  For a given coalgebra $c\colon C\to FC$,
  \Cref{algPartRefSetFun} computes behavioural equivalence.
\end{theorem}
\begin{proofappendix}{mainCorrectness}
  \textbf{Correctness:} We show that the property
  \[
    \text{for all clean states }x,y\in C\text{ with }p(x) = p(y)\text{ we have }
    F[p](c(x)) = F[p](c(y)).
    \tag*{\ensuremath{(\star)}}
    \label{algInv}
  \]
  holds throughout the execution of \Cref{algPartRefSetFun}:
  \begin{itemize}
  \item Initially, all states are marked dirty, so \ref{algInv} holds trivially.
  \item For every loop iteration for a block $B\in C/p$, let
    \begin{itemize}
    \item $p\colon C\to \N$ be the partition at the beginning of the loop iteration;
    \item $\phi \subseteq C$ be the clean states in $p$;
    \item $p'\colon C\to \N$ be the new partition (i.e.~the new value of $p$ after the splitting operation) after the loop iteration;
    \item $\phi' \subseteq C$ be the clean states after the loop iteration.
    \end{itemize}
    In other words, we assume
    \[
      \text{for all }x,y\in \phi\colon \quad p(x) = p(y) \text{ implies }F[p](c(x)) = F[p](c(y))
      \tag*{\ref{algInv}}
    \]
    and need to show that \ref{algInv} holds after the loop iteration, i.e.
    \begin{equation}
      \text{for all }x,y\in \phi'\colon \quad p'(x) = p'(y) \text{ implies }F[p'](c(x)) = F[p'](c(y))
      \tag*{(\ensuremath{\star'})}
      \label{algInvPrime}
    \end{equation}

    By \ref{algInv}, the composed map
    \(
      B \hookrightarrow C \xrightarrow{c} FC \xrightarrow{F[p]} F\N
    \)
    sends all clean states $x,y\in B\cap \phi$ to the same value, so it suffices
    to compute the signature $F[p](c(x))$ for one arbitrary clean state $x\in \BCone$.
    Since $\sig(x,p) = \sig(y,p)$ iff $F[p](c(x)) = F[p](c(y))$,
    the resulting partition $p'$ is then constructed by $\Split(B,A)$ such that
    \begin{align}
      p'(x) = p'(y)\quad\text{iff}\quad F[p](c(x)) = F[p](c(y)) \qquad\text{for all $x,y\in B$}.
      \label{pPrimeStable}
    \end{align}
    If a state $x\in C$ is clean at the end of the loop body, i.e.~$x\in \phi'$, this
    implies that $x$ has no successor $s$ in a block $B'\in \vec{B}_{new}$
    (otherwise $\MarkDirty(x)$ would have been called).
    The new blocks returned by $\Split$ indicate which states $s\in C$ were
    moved to a different block identifier: if $p(s) \neq p'(s)$, then
    $s\in B'$ for some $B' \in \vec{B}_{new}$. Combining these two observations
    yields that for all successors $s$ of $x$, we have $p(s) = p'(s)$, so by
    \Cref{sameSigForCleans}:
    \begin{align}
      F[p](c(x)) = F[p'](c(x)) \qquad\text{for all $x\in \phi'$}
      \label{cleanProperty}
    \end{align}
    For the final verification of \ref{algInvPrime}, consider clean states
    $x,y\in \phi'$ in the same block ($p'(x) = p'(y)$). In particular $x\in B$ iff $y\in B$, and we can show
    $F[p](c(x)) = F[p](c(y))$ by case distinction:
    \begin{itemize}
    \item If $x,y\in B$, then $F[p](c(x)) = F[p](c(y))$ by
    construction of $p'$ \eqref{pPrimeStable}.

    \item If $x,y\notin B$, then
    $p(x) = p(y)$ and so the states were clean
    before the current loop iteration, for which the invariant \ref{algInv}
    provides us with $F[p](c(x)) = F[p](c(y))$.
    \end{itemize}
    In any case, we have $F[p](c(x)) = F[p](c(y))$ and so
    by the observation for clean states \eqref{cleanProperty}
    \[
      F[p'](c(x))
      = F[p](c(x))
      = F[p](c(y))
      = F[p'](c(y))
    \]
    as desired, proving the invariant \ref{algInvPrime} after each loop iteration.
  \end{itemize}

  The invariant \ref{algInv} provides partial correctness:
  Whenever the algorithm terminates, the invariant \ref{algInv} shows that we
  have a well-defined map
  \[
    C/p \longrightarrow F(C/p)
  \]
  on the $p$-equivalence classes of $C$
  turning $p\colon C\to C/p$ into an $F$-coalgebra homomorphism.

  Thus, all states identified by $p$ are behaviourally equivalent. For the
  converse, one can show by induction over loop iterations that whenever two
  states $x,y\in C$ are behaviourally equivalent, then they remain identified by
  $p$.

  In total, upon termination, the returned partition $p$ precisely identifies
  the behaviourally equivalent states.
  Termination itself is clear because every finite set $|C|$ has only finitely
  many quotients.
\end{proofappendix}

\subsection{Complexity Analysis}
\label{sec:ComplexityAnalysis}
We structure the complexity analysis as a series of lemmas phrased in terms of the number of states $n = |C|$ and the total number of transitions $m$ defined by
\[
  m := \sum_{x\in C} |\pred(x)|
\]
As a first observation, we exploit that $\Split$ re-uses the block index for
the largest resulting block. Thus, whenever $x$ is moved to a block with a
different index, the new block has at most half the size of the old block,
leading to the logarithmic factor, by Hopcroft's trick:

\begin{lemma}
  \label{partLogNChange}
  A state is moved into a new block at most $\CO(\log n)$ times, that is, for
  every $x\in C$, the value of $p(x)$ in \Cref{algPartRefSetFun} changes at
  most $\lceil{\log_2 |C|}\rceil$ many times.
\end{lemma}
\begin{proofappendix}{partLogNChange}
  When a block gets split into sub-blocks, the old block is reused for the largest sub-block.
  Therefore, a newly created block is at most half the size of the old block.
  Formally, let $p_\old \colon C\to\N$ be the partition before an iteration of
  \Cref{algPartRefSetFun} and $p\colon C\to\N$ the partition after the
  iteration.  Then for all $x\in C$ we have
  \[
    p_\old (x) = p(x)
    \quad\text{or}\quad
    |\{x'\in C\mid p_\old(x) = p_\old(x')\}|
    ~~\ge~~ 2\cdot |\{x'\in C\mid p(x) = p(x')\}|
  \]
  In other words, each time a state is moved to a new block, the size of its containing block gets cut at least in half.
  Since the initial block has size $n$, the value of $p(x)$ can change at most
  $\lceil \log_2 n \rceil$ many times.
\end{proofappendix}

When a state is moved to a different block, all its predecessors are marked
dirty. If there are $m$ transitions in the system, and each state is moved to different block
at most $\log n$ times, then:
\begin{lemma}\label{markDirtyCount}
  $\MarkDirty$ is called at most $m\cdot \ceil{\log n} + n$ many times (including initialization).
\end{lemma}
\begin{proofappendix}{markDirtyCount}
  The initialization phase marks all the $n$ states of the coalgebra as dirty.

  Whenever $p(x)$ changes value, i.e.~$x$ is moved to another block, every
  $y\in \pred(x)$ is marked dirty. Using the bound from
  \autoref{partLogNChange}, we have that in the predecessor-loop, the total
  number of invocations of $\MarkDirty$ is bounded by
  \[
    \sum_{x\in C} (|\pred(x)|\cdot \log\ceil{n}) =
    (\sum_{x\in C} |\pred(x)|)\cdot \log\ceil{n} = m\cdot \log\ceil{n}
  \]
  leading to a total of at most $m\cdot \log\ceil{n} + n$ invocations.
\end{proofappendix}

In the actual implementation, we arrange the pointers in the initial partition
directly such that all states are marked dirty when the main loop is entered
for the first time.
The overall run time is dominated by the complexity of $\sig$ and $\pred$. Here,
we assume that $\sig$ always takes at least the time needed to write its return
value. On the other hand, we allow that $\pred$ returns a pre-computed array by
reference, taking only $\CO(1)$ time. The pre-computation of $\pred$ can be
done at the beginning of the algorithm by iterating over the entire coalgebra
once, e.g.~it can be done along with input parsing. This runs linear in the
overall size of the coalgebra, and thus is dominated by the complexity of the
algorithm:
\begin{proposition}
  \label{runTimeDominance}
  The run time complexity of \Cref{algPartRefSetFun} amounts to the time spent in
  $\sig$ and in $\pred$ plus $\CO(m\cdot \log n+n)$.
\end{proposition}
\begin{proofappendix}{runTimeDominance}
  We split the analysis in three sections: The initialization, the computation of the new partition, and the $\MarkDirty$ loop.
  \begin{itemize}
    \item The initialization takes $\CO(|C|)$. Since all states are marked dirty, the algorithm calls
      $\sig$ precisely $|C|$ times in  the first while-loop iteration. Hence, the initialization phase is
      dominated by the time spent in $\sig$.

    \item Extracting a block $B$ with dirty states takes $\CO(1)$, and computing
      the arrays of size $|\BD|$ and $|\BCone|$ is clearly dominated by the time spent in $\sig$.
      In particular, for every $x\in \BD$, the size of $sigs_\dirty(x)$ is bounded by the time $\sig(x,p)$.
      $\renumber'$ and the nested $\renumber$ run linearly in the total size of
      the signatures in $sigs_\dirty$ and $sigs_\clean$, whose contents were written by $\sig$.
      The invocation of $\Split$ runs in $\CO(|B_\dirty|)$, because the largest
      subblock of $B$ inherits the index of $B$; usually the subblock of clean
      states are the largest one and otherwise, the number of clean states is
      bounded by $|B_\dirty|$.

    \item $\MarkDirty$ runs in $\CO(1)$ so the time of the for-loops
      amounts to the number of iterations and the time spent in $\pred$. The
      \emph{outer} for-loop has at most $|B_\dirty$ iterations, because it iterates
      over all elements in the blocks returned by $\Split$.
      The \emph{inner} for-loop is bounded by the time and return value of
      $\pred$.
      If $\pred$ returns an array by reference in $\CO(1)$, then the inner
      for-loop amounts to the calls to $\MarkDirty$, contributing the extra
      $\CO(m\log n + n)$ time from \Cref{markDirtyCount}.
      \qedhere
  \end{itemize}
\end{proofappendix}

Thus, it remains to count how often the algorithm calls $\sig$.
Roughly, $\sig$ is called for every state that becomes dirty, so we can show:

\begin{theorem}\label{mainComplexity}
  The number of invocations of $\sig$ in \Cref{algPartRefSetFun} is bounded by
  $\CO(m \cdot \log n + n)$.
\end{theorem}
\begin{proofappendix}{mainComplexity}
  We show that $\sig$ is called at most
  \[
    2\cdot (m\ceil{\log n} + n)
  \]
  many times. There are two lines in the algorithm in which $\sig$ is called:
  \begin{enumerate}
    \item $\sig(x,p)$ is called for a \emph{dirty} state $x$. By \autoref{markDirtyCount}, this can happen at most
      $m\ceil{\log n} + n$ many times.
    \item $\sig(x,p)$ is called for a \emph{clean} state $x$. The bound for this is again
    $m\ceil{\log n} +n$, because if $\sig$ is called for a clean state $x\in
    \BCone \subseteq B$, then $B$ must have at least one dirty state. In
    particular, in every iteration for a block $B$, we have $|\BCone| \le
    |\BD|$.
  \end{enumerate}
  Hence, the overall number of invocations to $\sig$ is bounded by $2\cdot
  (m\ceil{\log n} + n)$ which is in $\CO(m\log n + n)$ as desired.
\end{proofappendix}

\begin{corollary}
  \label{corMain}
  If $\sig$ takes $f$ time, if $\pred$ runs in $\CO(1)$ (returning a reference) and $m\ge n$, then
  \Cref{algPartRefSetFun} computes behavioural equivalence in the input coalgebra in
  $\CO(f\cdot m\cdot \log n)$ time.
\end{corollary}
\begin{proofappendix}{corMain}
  By \Cref{runTimeDominance}, the time of $\sig$ dominates the overall run time. If $m\ge n$, the algorithm
  runs in $\CO(f\cdot (m\cdot \log n + n)) = \CO(f\cdot m\cdot \log n)$ time.
\end{proofappendix}

\begin{example}
  For $\Powf$-coalgebras, $\sig$ takes $\CO(k)$ time, if every state has at most $k$ successors.
  Then \Cref{algPartRefSetFun} minimizes $\Powf$-coalgebras in time $\CO(k\cdot m\cdot \log n)$.
  Note that $m\le k\cdot n$, so the complexity is also bounded by $\CO(k^2\cdot n\log n)$.
\end{example}

\subsection{Comparison to related work on the algorithmic level}

\newcommand{\algname}[1]{\textit{#1}}
\newcommand{\copar}[1]{\textit{CoPaR}}
\newcommand{\distr}[1]{\textit{DCPR}}
\newcommand{\mcrl}{\textit{mCRL2}}
\newcommand{\ours}[1]{\textit{Boa}}

We can classify partition refinement algorithms by their time complexity, and by the classes of functors they are applicable to.
For concrete system types, there are more algorithms than we can recall, so
instead, we focus on early representatives and on generic algorithms.

\subsubsection*{The Hopcroft line of work.}
One line of work originates in Hopcroft's 1971 work on DFA minimization \cite{Hopcroft71},
and continues with Kanellakis and Smolka's
\cite{KanellakisSmolka83,KanellakisS90} work on partition refinement for
transition systems running in $\CO(k^2 n\log n)$ where $k$ is the maximum
out-degree. It was a major achievement by Paige and Tarjan \cite{PaigeTarjan87}
to reduce the run time to $\CO(k n \log n)$ by counting transitions and storing these transition counters in a clever way,
which subsequently lead to a fruitful line of research on transition system minimization \cite{GaravelL22}.
This was generalized to coalgebras in Deifel, Dorsch, Milius, Schröder and Wißmann's work on \copar{},
which is applicable to a large class of functors satisfying their \emph{zippability} condition.
These algorithms keep track of a worklist of blocks with respect to which \emph{other} blocks still have to be split.
Our algorithm, by contrast, keeps track of a worklist of blocks that \emph{themselves} still potentially have to be split.
Although similar at first sight, they are fundamentally different: in the former, one is given a block,
and must determine how to split all the predecessor blocks,
whereas in our case one is given a block, which is then split based on its successors.

The advantage of the former class of algorithms is that they have optimal time complexity $\CO(kn \log n)$,
provided one can implement the special splitting procedure for the functor.
The additional memory needed for the transition counters is linear in $kn$.

Our algorithm, by contrast, has an extra factor of $k$,
but is applicable to all computable set-functors.
By investing this extra time-factor $k$, we reduce the memory consumption because we do not need to maintain transition counters or intermediate states like \copar{}.

A practical advantage of our algorithm is that \emph{one} recomputation of a block split can take into account the changes to \emph{all} the other blocks that happened since the recomputation.
The Hopcroft-\copar{} line of work, on the other hand, has to consider each change of the other blocks separately.
This advantage is of no help in the asymptotic complexity, because in the worst case only one other split happened each time,
and then our algorithm does in $\CO(k)$ what \copar{} can do in $\CO(1)$.
However, as we shall see in the benchmarks of \Cref{sec:Benchmarks}, in practice our algorithm outperforms \copar{} and \mcrl{},
even though our algorithm is applicable to a more general class of functors.

\subsubsection*{The Moore line of work.}
Another line of work originates in Moore's 1956 work on DFA minimization \cite{Moore},
which in retrospect is essentially the naive algorithm specialized to DFAs.
In this class, the most relevant for us is the algorithm by König and Küpper~\cite{KonigKupper14} for coalgebras,
and the distributed algorithm of Birkmann, Deifel, and Milius \cite{BDM22}.
Like our algorithm, algorithms in this class split a block based on its successors, and can be applied to general functors.
Unlike the Hopcroft-\copar{} line of work and our algorithm, the running time of these algorithms is $\CO(kn^2)$.

Another relevant algorithm in this class is the algorithm of Blom and Orzan \cite{BlomOrzan05} for transition systems.
Their main algorithm runs in time $\CO(kn^2)$, but in a side note they mention a variation of their algorithm that runs in $\CO(n \log n)$ \emph{iterations}.
They do not further analyse the time complexity or describe how to implement an iteration,
because the main focus of their paper is a distributed implementation of the $\CO(kn^2)$ algorithm,
and the $\CO(n \log n)$ variation precludes distributed implementation.
Out of all algorithms, Blom and Orzan's $\CO(n \log n)$ variation is the most similar to our algorithm,
in particular because their algorithm is in the Moore line of work, yet also re-uses the old block for the largest sub-block
(which is a feature that usually appears in the Hopcroft-\copar{} line of work).
However, their block splitting is different from ours and is only correct for
labelled transition systems but can not be easily applied to general functors $F$.

\section{Instances}
\label{sec:Instances}

We give a list of examples of instances that can be supported by our algorithm.
We start with the instances that were already previously supported by \copar{},
and then give examples of instances that were not previously supported by $n \log n$ algorithms.

\subsection{Instances also supported by \copar{}}

\subsubsection*{Products and coproducts}
The simplest instances are those built using the product $F \times G$ and
disjoint union $F + G$, or in general, signature functors for countable signatures $\Sigma$.
The binary encoding of an element of signature functor
$(\sigma,x_1,\ldots,x_k)\in \tilde\Sigma X$
starts with a specification of $\sigma$, followed by the concatenation of
encodings of the parameters $x_1,\ldots,x_k$. The functor implementation can
simply apply the substitution recursively to these elements $x_1,\ldots, x_k$,
without any further need for normalization.

\subsubsection*{Powerset}
The finite powerset functor $\Powf$ can be used to model transition systems as coalgebras.
In conjunction with products and coproducts, we can model nondeterministic
(tree) automata and labelled transition systems.
The binary encoding of an element $\{x_1,\ldots,x_k\}$ of the powerset functor,
is stored as a list of elements prefixed by its length.
The functor implementation can recursively apply the substitution to
the elements of a set $\{x_1,\ldots,x_k\}$,
and subsequently normalize by sorting the resulting elements and removing adjacent duplicates.

\subsubsection*{Monoid-valued functors}
The binary encoding of $\mu \in M^{(X)}$ (for a countable monoid $M$) is an
array of pairs $(x_i,\mu(x_i))$.
The binary encoding stores a list of these pairs prefixed by the length of the list.
The functor implementation recursively applies the substitution to the $x_i$,
and then sorts the pairs by the $x_i$ value,
and removes adjacent duplicate $x_i$ by summing up their associated monoid values $\mu(x_i)$.

\subsection{Instances not supported by \copar{}}

\subsubsection*{Composition of functors without intermediate states}

The requirement of zippability in the $m\log n$ algorithm~\cite{coparFM19} is
not closed under the composition of functors $F\circ G$. As a workaround, one
can introduce explicit intermediate states between $F$- and $G$-transitions.
This introduces potentially many more states into the coalgebra, which leads to
increased memory usage. Because our algorithm works for any computable functor,
it can instead use the composed functor directly,
without any pre-processing that splits each state of the automaton.
This is important for practical efficiency.

\subsection*{Monotone Modal Logics and Monotone Bisimulation}
When reasoning about game-theoretic settings~\cite{Parikh1985,Peleg87,Pauly2001}, the
arising modal logics have modal operators that talk about the ability of agents
to enforce properties in the future.
This leads to \emph{monotone modal logics} whose domain of reasoning are
\emph{monotone neighbourhood frames} and the canonical notion of equivalence is
\emph{monotone bisimulation}. It was shown by Hansen and
Kupke~\cite{HansenKupke04,HansenKupke04cmcs} that these are an instance of
coalgebras and coalgebraic behavioural equivalence for the monotone
neighbourhood functor. Instead of the original definition, it suffices for our
purposes to work with the following equivalent characterization:
\begin{notheorembrackets}
\begin{definition}[{\cite[Lem~3.3]{HansenKupke04}}]
  \label{defMonNeighbour}
  The monotone neighbourhood functor \linebreak
  ${\M\colon \Set\to\Set}$
  is given by
  \[
    \M X = \{ N\in \Powf\Powf X  \mid N\text{ \upshape upwards closed}\}
    \text{ and }
    \M(f\colon X\to Y)(N) = \closure{\{f[S]\mid S\in N\}}.
  \]
  where $\uparrow$ denotes upwards closure.
\end{definition}
\end{notheorembrackets}
\begin{proofappendix}[Details for]{defMonNeighbour}
  Usually $\N$ is defined to be a subfunctor of the double contravariant
    powerset functor $\M X \subseteq 2^{2^{X}}$, but it can also be defined
    using double covariant powerset $\Powf\Powf$ by explicitly defining upwards
    closure in the definition of maps $\M f\colon \M X\to \M Y$~\cite{HansenKupke04}.
\end{proofappendix}

Hence, in a coalgebra $c\colon C\to \M C$, the successor structure of a
state $x\in C$ is an upwards closed family of neighbourhoods $c(x)$.

To avoid redundancy, we do not keep the full neighbourhoods in memory, but only
the least elements in this family: given a family $N\in \M X$ for finite $X$,
we define the map
\[
  \atom_X\colon \Powf\Powf X \to \Powf\Powf X
  \qquad
  \atom_X(N) = \{S\in N\mid \nexists S'\in N: S'\subsetneqq S\}
\]
which transforms a monotone family into an antichain by taking the minimal
elements in the monotone family.

\begin{definition}
  \label{defMonImpl}
  We can implement $\M$-coalgebras as follows:
  For a coalgebra $c\colon C\to \M C$, keep for every
  state $x\in C$ an array of arrays representing $\atom_C(c(x)) \in \Powf\Powf X$.
  The predecessors of a state $y$ needs to be computed in advance and is given by
  \[
    \pred(y) = \{x\in C\mid y\in A\text{ for some }A\in \atom_C(c(x))\}.
  \]
  For the complexity analysis, we specify the out-degree as
  \[
    k := \max_{x\in C} \sum_{S\in \atom_C(c(x))} |S|.
  \]
  For the signature $\sig(x,p)$ of a state $x$ w.r.t.~$p\colon C\to \N$, do the following:
  \begin{enumerate}
  \item Compute $\Powf[\Powf [p]](t)$ for $t := \atom_C(c(x))$ by using the $\sig$-implementation of $\Powf$ first for each nested set and then on the outer set.
    This results in a new set of sets $t' := \Powf[\Powf [p]](t)$.
  \item For the normalization, iterate over all pairs $S,T\in t'$ and remove $T$
    if $S\subsetneqq T$.
    This step is not linear in the size of $t'$ but takes $\CO(k^2)$ time.
  \end{enumerate}
\end{definition}
\begin{proofappendix}[Details for]{defMonImpl}
  The only tricky part is why the normalization step in $\sig$ takes only
  $\CO(k^2)$ time.
  For each $S,T\in t'\in \Powf\Powf\N$, we compare $S\subsetneqq T$. Of course
  we have $\le k^2$ pairs $S,T$ and the comparison $S\subsetneqq T$ takes $k$
  steps, so we are clearly in $\CO(k^3)$. But actually we can give a tighter bound:
  We can address every $y\in S$ for every $S\in t'$ by an index $i$, using that
  we have just ordered $t'$ in the steps before. By the definition of $k$, each
  such index is smaller than $k$.
  Throughout checking $S\subsetneqq T$ for all $S,T\in t'$, we compare each
  pair of indices at most once. Hence, we perform at most $k^2$ comparisons.
\end{proofappendix}

For such a monotone neighbourhood frame $c\colon C\to \M C$, note that for
states $x\in C$, another state $y\in C$ might be contained in multiple sets
$S\in c(x)$. Still, the definition of $m$ in the complexity analysis is agnostic of this.

\begin{proposition}
  \label{propMonotone}
  For a monotone neighbourhood frame $c\colon C\to \M C$, let $k\in \N$ be such that $|\atom_C(c(x))| \le k$ for all $x\in C$. \Cref{algPartRefSetFun} computes monotone bisimilarity in $\CO(k^2\cdot m\log n)$ time.
\end{proposition}
\begin{proofappendix}{propMonotone}
  With $k\in \N$ denoting the bound on the out-degree, each call to $\sig$ takes $\CO(k^2)$ time.
  \Cref{algPartRefSetFun} calls $\sig$ $\CO(m\log n)$ many times, yielding $\CO(k^2\cdot m\log n)$ as desired.
\end{proofappendix}

\section{Benchmarks}
\label{sec:Benchmarks}

To evaluate the practical performance and memory usage of our algorithm,
we have implemented it in our tool \thetool{} \cite{artifact_boa}, written in Rust.
The user of \thetool{} can either use a composition of the built-in functors to describe their automaton type,
or implement their own automaton type by implementing the interface of \Cref{sec:coalginterface} in Rust.
The user may then input the data of their automaton using either a textual format akin to the representation in the ``Coalgebra'' row of \Cref{fig:coalgexamples},
or use \thetool{}'s more efficient and compact binary input format.

We test \thetool{} on the benchmark suite of Birkmann, Deifel and
Milius \cite{BDM22},
consisting of real-world benchmarks (fms \& wlan -- from the benchmark suite of the PRISM model checker \cite{PRISM}), and randomly generated benchmarks (wta -- weighted tree automata).
For the wta benchmarks,
the size of the first 5 was chosen to be maximal such that \copar{} \cite{coparFM19} uses 16GB of memory,
and the size of the 6th benchmark was chosen by Birkmann, Deifel and
Milius to demonstrate the scalability of their distributed algorithm.

\renewcommand\theadfont{\bfseries}

\newcommand{\tnodes}{\raisebox{.4ex}{\tiny{$\times 32$}}}
\newcommand{\tna}{--\ \ }
\newcommand{\tc}[2]{\multirow{#1}{*}{#2}}

\begin{table}[h!]
  \centering
\begin{tabular}{>{\bfseries}cccccccccc}
  \multicolumn{4}{l}{\thead{benchmark}} & \multicolumn{3}{l}{\thead{time (s)}} & \multicolumn{2}{c}{\thead{memory (MB)}} \\
  \toprule
  \thead{type} & \thead{n} & \thead{\% red} & \thead{m} & \thead{\copar{}} & \thead{\distr{}} & \thead{\ours{}} & \thead{\distr{}} & \thead{\ours{}} \\
  \toprule
  fms &           35910 &             0\% &          237120 &               4 &               2 &            0.02 &       13\tnodes &               6 \\
  fms &          152712 &             0\% &         1111482 &              17 &               8 &            0.10 &       62\tnodes &              20 \\
  fms &          537768 &             0\% &         4205670 &              68 &              26 &            0.40 &      163\tnodes &              72 \\
  fms &         1639440 &             0\% &        13552968 &             232 &              84 &            1.29 &      514\tnodes &             199 \\
  fms &         4459455 &             0\% &        38533968 &            \tna &             406 &            4.60 &     1690\tnodes &             557 \\
\midrule
 wlan &          248503 &            56\% &          437264 &              39 &             297 &            0.11 &       90\tnodes &              15 \\
 wlan &          607727 &            59\% &         1162573 &             105 &             855 &            0.30 &      147\tnodes &              38 \\
 wlan &         1632799 &            78\% &         3331976 &            \tna &            2960 &            0.81 &      379\tnodes &              92 \\
\midrule
wta$_5$(2) &           86852 &             0\% &        21713000 &             537 &              71 &            0.85 &      701\tnodes &             179 \\
wta$_4$(2) &           92491 &             0\% &        18498200 &             723 &              67 &            0.96 &      728\tnodes &             154 \\
wta$_3$(2) &          134207 &             0\% &        20131050 &             689 &             113 &            1.34 &      825\tnodes &             175 \\
wta$_2$(2) &          138000 &             0\% &        13800000 &             467 &             129 &            0.98 &      715\tnodes &             126 \\
wta$_1$(2) &          154863 &             0\% &         7743150 &             449 &             160 &            0.74 &      621\tnodes &              80 \\
wta$_3$(2) &         1300000 &             0\% &       195000000 &            \tna &            1377 &           22.58 &     7092\tnodes &            1647 \\
\midrule
wta$_5$(W) &           83431 &             0\% &        16686200 &             642 &              52 &            1.01 &      663\tnodes &             142 \\
wta$_4$(W) &           92615 &             0\% &        23153750 &             511 &              61 &            1.21 &      849\tnodes &             193 \\
wta$_3$(W) &           94425 &             0\% &        14163750 &             528 &              59 &            0.76 &      639\tnodes &             124 \\
wta$_2$(W) &          134082 &             0\% &        13408200 &             471 &              76 &            0.96 &      675\tnodes &             124 \\
wta$_1$(W) &          152107 &             0\% &         7605350 &             566 &              79 &            0.76 &      642\tnodes &              82 \\
wta$_3$(W) &          944250 &             0\% &       141637500 &            \tna &             675 &           15.18 &     6786\tnodes &            1231 \\
\midrule
wta$_5$(Z) &           92879 &             0\% &        18575800 &             463 &              56 &            0.67 &      754\tnodes &             161 \\
wta$_4$(Z) &           94451 &             0\% &        23612750 &             445 &              61 &            0.81 &      871\tnodes &             199 \\
wta$_3$(Z) &          100799 &             0\% &        15119850 &             391 &              64 &            0.62 &      628\tnodes &             135 \\
wta$_2$(Z) &          118084 &             0\% &        11808400 &             403 &              74 &            0.66 &      633\tnodes &             113 \\
wta$_1$(Z) &          156913 &             0\% &         7845650 &             438 &              82 &            0.68 &      677\tnodes &              93 \\
wta$_3$(Z) &         1007990 &             0\% &       151198500 &            \tna &             645 &           19.55 &     5644\tnodes &            1325 \\
\bottomrule
\end{tabular}
\caption{
  Time and memory usage comparison on the benchmarks of Birkmann, Deifel and
  Milius~\cite{BDM22}.
  The columns n, \%red, m give the number of states, the percentage of redundant states, and the number of edges, respectively.
  The results for \ours{} are an average of 10 runs.
  The results for \copar{} and \distr{} are those reported in Birkmann, Deifel and
  Milius \cite{BDM22}.
  The memory usage of \distr{} is per worker, indicated by $\times 32$ (for the 32 workers on the HPC cluster) \\
  The functors associated with the benchmarks are as follows:
   \textbf{fms:} $F(X) = \Q^{(X)}$,
   \textbf{wlan:} $F(X) = \N \times \Powf(\N \times \Dist(X))$,
   \textbf{wta$_r$(M):} $F(X) = M \times M^{(4 \times X^r)}$ where $r$ indicates the branching factor of the tree automaton,
      and $M=W$ is the monoid of 64-bit words with bitwise-or,
      $M=Z$ is the monoid of integers with addition,
      and $M=2$ is the monoid of booleans with logical-or.
}
\label{tab:benchmarks}
\end{table}

The benchmark results are given in \Cref{tab:benchmarks}.
The first columns list the type of benchmark and the size of the input coalgebra.
For the size, the column $n$ denotes the number of states and $m$ is the number
of edges as defined in \Cref{sec:ComplexityAnalysis}.
In the wlan benchmarks for CoPaR~\cite{coparFM19,WissmannEA2021}, the reported number of states
and eges also include intermediate states introduced by CoPaR in order to cope with functor
composition, a preprocessing step which we do not need in \ours{}, and thus are different from the numbers in \Cref{tab:benchmarks} here.

The three subsequent columns list the running time of \copar{}, \distr{}, and \ours{}.
The last two columns list the memory usage of \distr{} and \ours{}.
The benchmark results for \distr{} and \copar{} are those reported by Birkmann, Deifel and
Milius \cite{BDM22},
and were run on their high performance computing cluster with 32 workers on 8 nodes with two Xeon 2660v2 chips (10 cores per
chip + SMT) and 64GB RAM.
The memory usage of \distr{} is \emph{per worker}, indicated by the $\times 32$.

Execution times of \copar{} were taken using one node of the cluster.
Some entries for \copar{} are missing, indicating that it ran out of its 16GB of memory.
The benchmark results for our algorithm were obtained on a consumer setup: on one core of a 2.3GHz MacBook Pro 2019 with 32GB of memory.

A point to note is that compared to \copar{}, the distributed algorithm does best on the randomly generated benchmarks.
The distributed algorithm beats \copar{} in execution time by taking advantage of the large parallel compute power of the HPC cluster.
This comes at the cost of $\CO(n^2)$ worst case complexity, but randomly generated benchmarks are more or less the \emph{best case} for the distributed algorithm, and require only a very small constant number of iterations, so that the effective complexity is $\CO(n)$.
The real world benchmarks on the other hand, and especially the wlan benchmarks, need more iterations, which results in sequential \copar{} outperforming \distr{}.
In general, benchmarks with transition systems with long shortest path lengths will truly trigger the worst case of the $\CO(n^2)$ algorithm,
and can make its execution time infeasably long.
In summary, the benchmarks here are not chosen to be favourable to \copar{} and our algorithm, as they do not trigger the time complexity advantage to the full extent.

Nevertheless, our algorithm outperforms both \copar{} and \distr{} by a large margin. On the synthetic benchmarks (wta), roughly speaking, when \copar{} takes 10 minutes, \distr{} takes one minute, and our algorithm takes a second.
On the real-world wlan benchmark, the difference with \distr{} is greatest, with the largest benchmark requiring almost an hour on the HPC cluster for \distr{}, whereas our algorithm completes the benchmark in less than a second on a single thread.

Sequential \copar{} is unable to run the largest wta benchmarks, because it requires more memory than the 16GB limit.
The distributed algorithm is able to spread the required memory usage among 32 workers, thus staying under the 16GB limit per worker.
Our algorithm uses sufficiently less memory to be able to run all benchmarks on a single machine. In fact, it uses significantly less memory than \distr{} uses \emph{per worker}.
There are several reasons for this:
\begin{itemize}
  \item Our algorithm does not require large hash tables.
  \item Our algorithm uses an binary representation with simple in-memory dictionary compression.
  \item We operate directly on the composed functor instead of splitting states into pieces.
\end{itemize}
Even the largest benchmarks stay far away from the 16GB memory limit.
We are thus able to minimize large coalgebraic transition systems on cheap, consumer grade hardware.

To assess the cost of genericity, we also compare with \mcrl{}, a full toolset
for the verification of concurrent systems. Among many other tasks, \mcrl{}
also supports minimization of transition systems by strong bisimilarity as part
of the \texttt{ltsconvert}
command\footnote{\url{https://www.mcrl2.org/web/user_manual/tools/release/ltsconvert.html}}
and even implements multiple algorithms for that, out of which the algorithm by Jansen \etal~\cite{JansenGKW20} turned out to be the fastest.
For benchmarking, we ran its implementation in \mcrl{} and compared the fastest with the run time of
$\ours{}$. As input files, we used the \emph{very large transition systems}
(VLTS) benchmark suite\footnote{\url{https://cadp.inria.fr/resources/vlts/}}.
Unfortunately, the benchmark suite is not available online in an open format,
so the files were converted with the CADP tool to the plain text \texttt{.aut}
format, supported by \mcrl{} and our tool. The results are shown in \autoref{tab:benchmarksmcrl}.
The benchmark consists of two series of input files, \textit{cwi} and \textit{vasy}, whose file sizes ranged from a few KB to hundreds of MB (biggest vasy ws 145MB in zipped format and biggest cwi was 630MB zipped).
Surprisingly, $\ours{}$ is significantly faster than the bisimilarity minimization implemented in \mcrl{}. On all input files, \mcrl{} and \ours{} agreed
on the size of the resulting partition, giving confidence in the correctness of the computed partition.
It should be noted that \mcrl{} supports a wide range of bisimilarity notions (e.g.~branching bisimilarity), which our algorithm can not cover.
%

%
%
%
%
%
%
%
%
%
%
%
%
%
%
%
%
%

%
%
%
%
%
%
%
%
%
%
%
%
%
%
%
%
%
%

\begin{table}[h!]
  \centering
\begin{tabular}{>{\bfseries}cccccccc}
  \multicolumn{4}{l}{\thead{benchmark}} & \multicolumn{2}{l}{\thead{time (s)}} & \multicolumn{2}{c}{\thead{memory (MB)}} \\
  \toprule
  \thead{type} & \thead{n} & \thead{\% red} & \thead{m} & \thead{\mcrl{}} &   \thead{\ours{}} &        \thead{\mcrl{}} &     \thead{\ours{}} \\
  \toprule
  cwi &          142472 &            97\% &          925429 &            0.85 &            0.08 &              99 &              15 \\
  cwi &          214202 &            63\% &          684419 &            0.63 &            0.15 &             111 &              16 \\
  cwi &          371804 &            90\% &          641565 &            0.38 &            0.11 &              95 &              22 \\
  cwi &          566640 &            97\% &         3984157 &            6.19 &            0.44 &             414 &              60 \\
  cwi &         2165446 &            98\% &         8723465 &           10.72 &            1.52 &             978 &             166 \\
  cwi &         2416632 &            96\% &        17605592 &           14.87 &            1.56 &            1780 &             247 \\
  cwi &         7838608 &            87\% &        59101007 &          231.08 &           17.43 &            5777 &             816 \\
  cwi &        33949609 &            99\% &       165318222 &          312.11 &           35.41 &           16698 &            2809 \\
\midrule
 vasy &           52268 &            84\% &          318126 &            0.31 &            0.04 &              48 &               7 \\
 vasy &           65537 &             0\% &         2621480 &            6.62 &            0.14 &             553 &              28 \\
 vasy &           66929 &             0\% &         1302664 &            2.56 &            0.08 &             275 &              18 \\
 vasy &           69754 &             0\% &          520633 &            0.93 &            0.04 &             128 &              11 \\
 vasy &           83436 &             0\% &          325584 &            0.38 &            0.04 &              86 &              10 \\
 vasy &          116456 &             0\% &          368569 &            0.47 &            0.06 &             105 &              15 \\
 vasy &          164865 &            99\% &         1619204 &            1.92 &            0.23 &             162 &              22 \\
 vasy &          166464 &            49\% &          651168 &            0.81 &            0.08 &             116 &              16 \\
 vasy &          386496 &            99\% &         1171872 &            0.67 &            0.08 &             133 &              28 \\
 vasy &          574057 &            99\% &        13561040 &           18.84 &            2.41 &            1277 &             141 \\
 vasy &          720247 &            99\% &          390999 &            0.38 &            0.05 &              88 &              31 \\
 vasy &         1112490 &            99\% &         5290860 &            8.86 &            0.78 &             579 &              93 \\
 vasy &         2581374 &             0\% &        11442382 &           31.95 &            2.30 &            2691 &             285 \\
 vasy &         4220790 &            67\% &        13944372 &           31.82 &            2.87 &            2293 &             311 \\
 vasy &         4338672 &            40\% &        15666588 &           34.89 &            3.12 &            3160 &             372 \\
 vasy &         6020550 &            99\% &        19353474 &           34.91 &            4.11 &            2124 &             534 \\
 vasy &         6120718 &            99\% &        11031292 &           15.56 &            2.37 &            1297 &             325 \\
 vasy &         8082905 &            99\% &        42933110 &           72.45 &            3.79 &            4313 &             719 \\
 vasy &        11026932 &            91\% &        24660513 &           60.57 &            6.26 &            2768 &             661 \\
 vasy &        12323703 &            91\% &        27667803 &           63.49 &            8.16 &            3103 &             740 \\
\bottomrule
\end{tabular}
\caption{
  Time and memory usage comparison on the VLTS benchmark suite (for space reasons, we have excluded the very short running benchmarks).
  The columns n, \%red, m give the number of states, the percentage of redundant states, and the number of edges, respectively.
  The results are an average of 10 runs.
  For \mcrl{}, the default \texttt{bisim} option was used, which runs the JGKW algorithm \cite{JansenGKW20}.
}
\vspace{-0.5cm}
\label{tab:benchmarksmcrl}
\end{table}

\section{Conclusions and Future Work}
\label{sec:Conclusion}
The coalgebraic approach enables generic tools for automata minimization,
applying to different types of input automata.
With our coalgebraic partition refinement algorithm,
implemented in our tool \thetool{},
we reduce the time and memory use compared to previous work.
This comes at the cost of an extra factor of $k$ (the outdegree of a state) in the time-complexity compared to asymptotically optimal algorithms.
Though our asymptotic complexity is not as good as the asymptotically fastest but less generic algorithms,
the evaluation shows the efficiency of our algorithm.

We wish to expand the supported system equivalence notions. So far, our algorithm is applicable to functors on \Set. More advanced
equivalence and bisimilarity notions such as trace
equivalence~\cite{SILVA2011291,HasuoJS07}, branching bisimulations, and others from the
linear-time-branching spectrum~\cite{Glabbeek01}, can be understood
coalgebraically using graded
monads~\cite{DorschMS19,MiliusPS15}, corresponding to changing the base
category of the functor from \Set{} to, for example, the Eilenberg-Moore~\cite{SilvaBBR13} or
Kleisli~\cite{HasuoJS07} category of a monad.
For branching bisimulation,
efficient algorithms exist~\cite{JansenGKW20,GrooteV90}, whose ideas might embed into our framework.
We conjecture that it is possible to adapt the algorithm to nominal
sets, in order to minimize (orbit-)finite coalgebras there~\cite{KozenEA15,msw16,skmw17,SuppSet23}.

Up-to techniques provide another successful line of research for deciding
bisimilarity. Bonchi and Pous~\cite{BonchiP13} provide a construction
for deciding bismilarity of two particular states of interest, where the
transition structure is unfolded lazily while the reasoning evolves. By
computing the partitions in a similarly lazy way, performance of our minimization algorithm can hopefully be improved even further.

\begin{acks}
  We thank
  Hans-Peter Deifel,
  Stefan Milius,
  Jurriaan Rot,
  Hubert Garavel,
  Sebastian Junges,
  Marck van der Vegt,
  Joost-Pieter Katoen,
  and Frits Vaandrager
   for helpful discussions and the anonymous referees for their valuable feedback for improving the paper.
  Thorsten Wißmann was supported by the NWO TOP project 612.001.852.
\end{acks}

\label{maintextend}
\bibliographystyle{ACM-Reference-Format}
\bibliography{refs.bib}


\providecommand{\noopsort}[1]{}
\begin{thebibliography}{53}


\ifx \showCODEN    \undefined \def \showCODEN     #1{\unskip}     \fi
\ifx \showDOI      \undefined \def \showDOI       #1{#1}\fi
\ifx \showISBNx    \undefined \def \showISBNx     #1{\unskip}     \fi
\ifx \showISBNxiii \undefined \def \showISBNxiii  #1{\unskip}     \fi
\ifx \showISSN     \undefined \def \showISSN      #1{\unskip}     \fi
\ifx \showLCCN     \undefined \def \showLCCN      #1{\unskip}     \fi
\ifx \shownote     \undefined \def \shownote      #1{#1}          \fi
\ifx \showarticletitle \undefined \def \showarticletitle #1{#1}   \fi
\ifx \showURL      \undefined \def \showURL       {\relax}        \fi
\providecommand\bibfield[2]{#2}
\providecommand\bibinfo[2]{#2}
\providecommand\natexlab[1]{#1}
\providecommand\showeprint[2][]{arXiv:#2}

\bibitem[Aho et~al\mbox{.}(1974)]%
        {AHU74}
\bibfield{author}{\bibinfo{person}{Alfred~V. Aho}, \bibinfo{person}{John~E.
  Hopcroft}, {and} \bibinfo{person}{Jeffrey~D. Ullman}.}
  \bibinfo{year}{1974}\natexlab{}.
\newblock \bibinfo{booktitle}{\emph{{The Design and Analysis of Computer
  Algorithms}}}.
\newblock \bibinfo{publisher}{Addison-Wesley}, \bibinfo{address}{Reading,
  Mass.}
\newblock


\bibitem[Baier et~al\mbox{.}(2000)]%
        {BaierEM00}
\bibfield{author}{\bibinfo{person}{Christel Baier}, \bibinfo{person}{Bettina
  Engelen}, {and} \bibinfo{person}{Mila Majster{-}Cederbaum}.}
  \bibinfo{year}{2000}\natexlab{}.
\newblock \showarticletitle{Deciding Bisimilarity and Similarity for
  Probabilistic Processes}.
\newblock \bibinfo{journal}{\emph{J.\ Comput.\ Syst.\ Sci.}}
  \bibinfo{volume}{60} (\bibinfo{year}{2000}), \bibinfo{pages}{187--231}.
\newblock


\bibitem[Baier and Katoen(2008)]%
        {BaierKatoen08}
\bibfield{author}{\bibinfo{person}{Christel Baier} {and}
  \bibinfo{person}{Joost{-}Pieter Katoen}.} \bibinfo{year}{2008}\natexlab{}.
\newblock \bibinfo{booktitle}{\emph{Principles of model checking}}.
\newblock \bibinfo{publisher}{{MIT} Press}.
\newblock
\showISBNx{978-0-262-02649-9}


\bibitem[Bartels et~al\mbox{.}(2003)]%
        {BARTELS200357}
\bibfield{author}{\bibinfo{person}{Falk Bartels}, \bibinfo{person}{Ana
  Sokolova}, {and} \bibinfo{person}{Erik de Vink}.}
  \bibinfo{year}{2003}\natexlab{}.
\newblock \showarticletitle{A hierarchy of probabilistic system types}. In
  \bibinfo{booktitle}{\emph{Coagebraic Methods in Computer Science, CMCS 2003}}
  \emph{(\bibinfo{series}{ENTCS}, Vol.~\bibinfo{volume}{82})}.
  \bibinfo{publisher}{Elsevier}, \bibinfo{pages}{57 -- 75}.
\newblock
\showISSN{1571-0661}


\bibitem[Birkmann et~al\mbox{.}(2022)]%
        {BDM22}
\bibfield{author}{\bibinfo{person}{Fabian Birkmann},
  \bibinfo{person}{Hans{-}Peter Deifel}, {and} \bibinfo{person}{Stefan
  Milius}.} \bibinfo{year}{2022}\natexlab{}.
\newblock \showarticletitle{Distributed Coalgebraic Partition Refinement}. In
  \bibinfo{booktitle}{\emph{Tools and Algorithms for the Construction and
  Analysis of Systems - 28th International Conference, {TACAS} 2022,
  Proceedings, Part {II}}} \emph{(\bibinfo{series}{LNCS},
  Vol.~\bibinfo{volume}{13244})}, \bibfield{editor}{\bibinfo{person}{Dana
  Fisman} {and} \bibinfo{person}{Grigore Rosu}} (Eds.).
  \bibinfo{publisher}{Springer}, \bibinfo{pages}{159--177}.
\newblock
\urldef\tempurl%
\url{https://doi.org/10.1007/978-3-030-99527-0\_9}
\showDOI{\tempurl}


\bibitem[Björklund et~al\mbox{.}(2007)]%
        {HoegbergEA07}
\bibfield{author}{\bibinfo{person}{Johanna~(H{\"{o}}gberg) Björklund},
  \bibinfo{person}{Andreas Maletti}, {and} \bibinfo{person}{Jonathan May}.}
  \bibinfo{year}{2007}\natexlab{}.
\newblock \showarticletitle{Bisimulation Minimisation for Weighted Tree
  Automata}. In \bibinfo{booktitle}{\emph{Developments in Language Theory,
  {DLT} 2007}} \emph{(\bibinfo{series}{LNCS}, Vol.~\bibinfo{volume}{4588})}.
  \bibinfo{publisher}{Springer}, \bibinfo{pages}{229--241}.
\newblock
\showISBNx{978-3-540-73207-5}


\bibitem[Björklund et~al\mbox{.}(2009)]%
        {HoegbergEA09}
\bibfield{author}{\bibinfo{person}{Johanna~(H\"{o}gberg) Björklund},
  \bibinfo{person}{Andreas Maletti}, {and} \bibinfo{person}{Jonathan May}.}
  \bibinfo{year}{2009}\natexlab{}.
\newblock \showarticletitle{Backward and forward bisimulation minimization of
  tree automata}.
\newblock \bibinfo{journal}{\emph{Theor.\ Comput.\ Sci.}}
  \bibinfo{volume}{410} (\bibinfo{year}{2009}), \bibinfo{pages}{3539--3552}.
\newblock


\bibitem[Blom and Orzan(2005)]%
        {BlomOrzan05}
\bibfield{author}{\bibinfo{person}{Stefan Blom} {and} \bibinfo{person}{Simona
  Orzan}.} \bibinfo{year}{2005}\natexlab{}.
\newblock \showarticletitle{Distributed state space minimization}.
\newblock \bibinfo{journal}{\emph{International Journal on Software Tools for
  Technology Transfer}} \bibinfo{volume}{7}, \bibinfo{number}{3}
  (\bibinfo{date}{June} \bibinfo{year}{2005}), \bibinfo{pages}{280--291}.
\newblock
\urldef\tempurl%
\url{https://doi.org/10.1007/s10009-004-0185-2}
\showDOI{\tempurl}


\bibitem[Bonchi and Pous(2013)]%
        {BonchiP13}
\bibfield{author}{\bibinfo{person}{Filippo Bonchi} {and}
  \bibinfo{person}{Damien Pous}.} \bibinfo{year}{2013}\natexlab{}.
\newblock \showarticletitle{Checking {NFA} equivalence with bisimulations up to
  congruence}. In \bibinfo{booktitle}{\emph{The 40th Annual {ACM}
  {SIGPLAN-SIGACT} Symposium on Principles of Programming Languages, {POPL}
  '13}}, \bibfield{editor}{\bibinfo{person}{Roberto Giacobazzi} {and}
  \bibinfo{person}{Radhia Cousot}} (Eds.). \bibinfo{publisher}{{ACM}},
  \bibinfo{pages}{457--468}.
\newblock
\urldef\tempurl%
\url{https://doi.org/10.1145/2429069.2429124}
\showDOI{\tempurl}


\bibitem[Deifel et~al\mbox{.}(2019)]%
        {coparFM19}
\bibfield{author}{\bibinfo{person}{Hans-Peter Deifel}, \bibinfo{person}{Stefan
  Milius}, \bibinfo{person}{Lutz Schr{\"o}der}, {and} \bibinfo{person}{Thorsten
  Wi{\ss}mann}.} \bibinfo{year}{2019}\natexlab{}.
\newblock \showarticletitle{Generic Partition Refinement and Weighted Tree
  Automata}. In \bibinfo{booktitle}{\emph{Formal Methods -- The Next 30 Years,
  Proc.~3rd World Congress on Formal Methods (FM 2019)}}
  \emph{(\bibinfo{series}{LNCS}, Vol.~\bibinfo{volume}{11800})}.
  \bibinfo{publisher}{Springer}, \bibinfo{pages}{280--297}.
\newblock
\showISBNx{978-3-030-30942-8}


\bibitem[Dorsch et~al\mbox{.}(2019)]%
        {DorschMS19}
\bibfield{author}{\bibinfo{person}{Ulrich Dorsch}, \bibinfo{person}{Stefan
  Milius}, {and} \bibinfo{person}{Lutz Schr{\"{o}}der}.}
  \bibinfo{year}{2019}\natexlab{}.
\newblock \showarticletitle{Graded Monads and Graded Logics for the Linear Time
  - Branching Time Spectrum}. In \bibinfo{booktitle}{\emph{30th International
  Conference on Concurrency Theory, {CONCUR} 2019}}
  \emph{(\bibinfo{series}{LIPIcs}, Vol.~\bibinfo{volume}{140})},
  \bibfield{editor}{\bibinfo{person}{Wan~J. Fokkink} {and} \bibinfo{person}{Rob
  van Glabbeek}} (Eds.). \bibinfo{publisher}{Schloss Dagstuhl - Leibniz-Zentrum
  f{\"{u}}r Informatik}, \bibinfo{pages}{36:1--36:16}.
\newblock
\urldef\tempurl%
\url{https://doi.org/10.4230/LIPIcs.CONCUR.2019.36}
\showDOI{\tempurl}


\bibitem[Dorsch et~al\mbox{.}(2017)]%
        {DorschEA17}
\bibfield{author}{\bibinfo{person}{Ulrich Dorsch}, \bibinfo{person}{Stefan
  Milius}, \bibinfo{person}{Lutz Schr{\"o}der}, {and} \bibinfo{person}{Thorsten
  Wi{\ss}mann}.} \bibinfo{year}{2017}\natexlab{}.
\newblock \showarticletitle{Efficient Coalgebraic Partition Refinement}. In
  \bibinfo{booktitle}{\emph{Proc.~28th International Conference on Concurrency
  Theory (CONCUR 2017)}} \emph{(\bibinfo{series}{LIPIcs})}.
  \bibinfo{publisher}{Schloss Dagstuhl - Leibniz-Zentrum fuer Informatik}.
\newblock


\bibitem[Garavel and Lang(2022)]%
        {GaravelL22}
\bibfield{author}{\bibinfo{person}{Hubert Garavel} {and}
  \bibinfo{person}{Fr{\'{e}}d{\'{e}}ric Lang}.}
  \bibinfo{year}{2022}\natexlab{}.
\newblock \showarticletitle{Equivalence Checking 40 Years After: {A} Review of
  Bisimulation Tools} \emph{(\bibinfo{series}{Lecture Notes in Computer
  Science})}.
\newblock
\urldef\tempurl%
\url{https://doi.org/10.1007/978-3-031-15629-8\_13}
\showDOI{\tempurl}


\bibitem[Groote and Vaandrager(1990)]%
        {GrooteV90}
\bibfield{author}{\bibinfo{person}{Jan~Friso Groote} {and}
  \bibinfo{person}{Frits~W. Vaandrager}.} \bibinfo{year}{1990}\natexlab{}.
\newblock \showarticletitle{An Efficient Algorithm for Branching Bisimulation
  and Stuttering Equivalence}. In \bibinfo{booktitle}{\emph{Automata, Languages
  and Programming, 17th International Colloquium, ICALP90, Warwick University,
  England, UK, July 16-20, 1990, Proceedings}} \emph{(\bibinfo{series}{Lecture
  Notes in Computer Science}, Vol.~\bibinfo{volume}{443})},
  \bibfield{editor}{\bibinfo{person}{Mike Paterson}} (Ed.).
  \bibinfo{publisher}{Springer}, \bibinfo{pages}{626--638}.
\newblock
\urldef\tempurl%
\url{https://doi.org/10.1007/BFb0032063}
\showDOI{\tempurl}


\bibitem[Groote et~al\mbox{.}(2018)]%
        {GrooteEA18}
\bibfield{author}{\bibinfo{person}{Jan~Friso Groote},
  \bibinfo{person}{Jao~Rivera Verduzco}, {and} \bibinfo{person}{Erik~P. de
  Vink}.} \bibinfo{year}{2018}\natexlab{}.
\newblock \showarticletitle{An Efficient Algorithm to Determine Probabilistic
  Bisimulation}.
\newblock \bibinfo{journal}{\emph{Algorithms}} \bibinfo{volume}{11},
  \bibinfo{number}{9} (\bibinfo{year}{2018}), \bibinfo{pages}{131}.
\newblock


\bibitem[Gumm and Schr{\"{o}}der(2001)]%
        {GummS01}
\bibfield{author}{\bibinfo{person}{H.~Peter Gumm} {and} \bibinfo{person}{Tobias
  Schr{\"{o}}der}.} \bibinfo{year}{2001}\natexlab{}.
\newblock \showarticletitle{Monoid-labeled transition systems}. In
  \bibinfo{booktitle}{\emph{Coalgebraic Methods in Computer Science, {CMCS}
  2001}} \emph{(\bibinfo{series}{ENTCS}, Vol.~\bibinfo{volume}{44(1)})}.
  \bibinfo{publisher}{Elsevier}, \bibinfo{pages}{185--204}.
\newblock


\bibitem[Hansen and Kupke(2004a)]%
        {HansenKupke04}
\bibfield{author}{\bibinfo{person}{Helle~Hvid Hansen} {and}
  \bibinfo{person}{Clemens Kupke}.} \bibinfo{year}{2004}\natexlab{a}.
\newblock \showarticletitle{A Coalgebraic Perspective on Monotone Modal Logic}.
\newblock \bibinfo{journal}{\emph{Electron. Notes Theor. Comput. Sci.}}
  \bibinfo{volume}{106} (\bibinfo{date}{December} \bibinfo{year}{2004}),
  \bibinfo{pages}{121--143}.
\newblock
\showISSN{1571-0661}
\urldef\tempurl%
\url{https://doi.org/10.1016/j.entcs.2004.02.028}
\showDOI{\tempurl}


\bibitem[Hansen and Kupke(2004b)]%
        {HansenKupke04cmcs}
\bibfield{author}{\bibinfo{person}{Helle~Hvid Hansen} {and}
  \bibinfo{person}{Clemens Kupke}.} \bibinfo{year}{2004}\natexlab{b}.
\newblock \showarticletitle{A Coalgebraic Perspective on Monotone Modal Logic}.
  In \bibinfo{booktitle}{\emph{Proceedings of the Workshop on Coalgebraic
  Methods in Computer Science, {CMCS}}} \emph{(\bibinfo{series}{Electronic
  Notes in Theoretical Computer Science}, Vol.~\bibinfo{volume}{106})},
  \bibfield{editor}{\bibinfo{person}{Jir{\'{\i}} Ad{\'{a}}mek} {and}
  \bibinfo{person}{Stefan Milius}} (Eds.). \bibinfo{publisher}{Elsevier},
  \bibinfo{pages}{121--143}.
\newblock
\urldef\tempurl%
\url{https://doi.org/10.1016/j.entcs.2004.02.028}
\showDOI{\tempurl}


\bibitem[Hasuo et~al\mbox{.}(2007)]%
        {HasuoJS07}
\bibfield{author}{\bibinfo{person}{Ichiro Hasuo}, \bibinfo{person}{Bart
  Jacobs}, {and} \bibinfo{person}{Ana Sokolova}.}
  \bibinfo{year}{2007}\natexlab{}.
\newblock \showarticletitle{Generic Trace Semantics via Coinduction}.
\newblock \bibinfo{journal}{\emph{Log. Methods Comput. Sci.}}
  \bibinfo{volume}{3}, \bibinfo{number}{4} (\bibinfo{year}{2007}).
\newblock
\urldef\tempurl%
\url{https://doi.org/10.2168/LMCS-3(4:11)2007}
\showDOI{\tempurl}


\bibitem[Hopcroft(1971)]%
        {Hopcroft71}
\bibfield{author}{\bibinfo{person}{John Hopcroft}.}
  \bibinfo{year}{1971}\natexlab{}.
\newblock \showarticletitle{An $n \log n$ algorithm for minimizing states in a
  finite automaton}. In \bibinfo{booktitle}{\emph{Theory of Machines and
  Computations}}. \bibinfo{publisher}{Academic Press},
  \bibinfo{pages}{189--196}.
\newblock


\bibitem[Jansen et~al\mbox{.}(2020)]%
        {JansenGKW20}
\bibfield{author}{\bibinfo{person}{David~N. Jansen}, \bibinfo{person}{Jan~Friso
  Groote}, \bibinfo{person}{Jeroen J.~A. Keiren}, {and} \bibinfo{person}{Anton
  Wijs}.} \bibinfo{year}{2020}\natexlab{}.
\newblock \showarticletitle{An O(m log n) algorithm for branching bisimilarity
  on labelled transition systems}. In \bibinfo{booktitle}{\emph{Tools and
  Algorithms for the Construction and Analysis of Systems - 26th International
  Conference, {TACAS} 2020, Held as Part of the European Joint Conferences on
  Theory and Practice of Software, {ETAPS} 2020, Dublin, Ireland, April 25-30,
  2020, Proceedings, Part {II}}} \emph{(\bibinfo{series}{Lecture Notes in
  Computer Science}, Vol.~\bibinfo{volume}{12079})},
  \bibfield{editor}{\bibinfo{person}{Armin Biere} {and} \bibinfo{person}{David
  Parker}} (Eds.). \bibinfo{publisher}{Springer}, \bibinfo{pages}{3--20}.
\newblock
\urldef\tempurl%
\url{https://doi.org/10.1007/978-3-030-45237-7\_1}
\showDOI{\tempurl}


\bibitem[Jules~Jacobs(2022)]%
        {artifact_boa}
\bibfield{author}{\bibinfo{person}{Thorsten~Wissmann Jules~Jacobs}.}
  \bibinfo{year}{2022}\natexlab{}.
\newblock \bibinfo{title}{Boa: binary coalgebraic partition refinement}.
\newblock
\newblock
\urldef\tempurl%
\url{https://doi.org/10.5281/zenodo.7150706}
\showDOI{\tempurl}
\newblock
\shownote{The most recent version is at
  \url{https://github.com/julesjacobs/boa}.}.


\bibitem[Kanellakis and Smolka(1983)]%
        {KanellakisSmolka83}
\bibfield{author}{\bibinfo{person}{Paris~C. Kanellakis} {and}
  \bibinfo{person}{Scott~A. Smolka}.} \bibinfo{year}{1983}\natexlab{}.
\newblock \showarticletitle{{CCS} Expressions, Finite State Processes, and
  Three Problems of Equivalence}. In \bibinfo{booktitle}{\emph{Proceedings of
  the Second Annual ACM Symposium on Principles of Distributed Computing}}
  (Montreal, Quebec, Canada) \emph{(\bibinfo{series}{PODC '83})}.
  \bibinfo{publisher}{ACM}, \bibinfo{pages}{228--240}.
\newblock
\showISBNx{0-89791-110-5}


\bibitem[Kanellakis and Smolka(1990)]%
        {KanellakisS90}
\bibfield{author}{\bibinfo{person}{Paris~C. Kanellakis} {and}
  \bibinfo{person}{Scott~A. Smolka}.} \bibinfo{year}{1990}\natexlab{}.
\newblock \showarticletitle{{CCS} Expressions, Finite State Processes, and
  Three Problems of Equivalence}.
\newblock \bibinfo{journal}{\emph{Inf. Comput.}} \bibinfo{volume}{86},
  \bibinfo{number}{1} (\bibinfo{year}{1990}), \bibinfo{pages}{43--68}.
\newblock


\bibitem[Katoen et~al\mbox{.}(2007)]%
        {KatoenEA07}
\bibfield{author}{\bibinfo{person}{Joost{-}Pieter Katoen}, \bibinfo{person}{Tim
  Kemna}, \bibinfo{person}{Ivan Zapreev}, {and} \bibinfo{person}{David
  Jansen}.} \bibinfo{year}{2007}\natexlab{}.
\newblock \showarticletitle{Bisimulation Minimisation Mostly Speeds Up
  Probabilistic Model Checking}. In \bibinfo{booktitle}{\emph{Tools and
  Algorithms for the Construction and Analysis of Systems, {TACAS} 2007}}
  \emph{(\bibinfo{series}{LNCS}, Vol.~\bibinfo{volume}{4424})}.
  \bibinfo{publisher}{Springer}, \bibinfo{pages}{87--101}.
\newblock
\showISBNx{978-3-540-71208-4}


\bibitem[Klin(2009)]%
        {Klin09}
\bibfield{author}{\bibinfo{person}{Bartek Klin}.}
  \bibinfo{year}{2009}\natexlab{}.
\newblock \showarticletitle{Structural Operational Semantics for Weighted
  Transition Systems}. In \bibinfo{booktitle}{\emph{Semantics and Algebraic
  Specification: Essays Dedicated to Peter D.~Mosses on the Occasion of His
  60th Birthday}} \emph{(\bibinfo{series}{LNCS}, Vol.~\bibinfo{volume}{5700})},
  \bibfield{editor}{\bibinfo{person}{Jens Palsberg}} (Ed.).
  \bibinfo{publisher}{Springer}, \bibinfo{pages}{121--139}.
\newblock


\bibitem[K{\"{o}}nig and K{\"{u}}pper(2014)]%
        {KonigKupper14}
\bibfield{author}{\bibinfo{person}{Barbara K{\"{o}}nig} {and}
  \bibinfo{person}{Sebastian K{\"{u}}pper}.} \bibinfo{year}{2014}\natexlab{}.
\newblock \showarticletitle{Generic Partition Refinement Algorithms for
  Coalgebras and an Instantiation to Weighted Automata}. In
  \bibinfo{booktitle}{\emph{Theoretical Computer Science, IFIP TCS 2014}}
  \emph{(\bibinfo{series}{LNCS}, Vol.~\bibinfo{volume}{8705})}.
  \bibinfo{publisher}{Springer}, \bibinfo{pages}{311--325}.
\newblock
\showISBNx{978-3-662-44601-0}


\bibitem[Kozen et~al\mbox{.}(2015)]%
        {KozenEA15}
\bibfield{author}{\bibinfo{person}{Dexter Kozen}, \bibinfo{person}{Konstantinos
  Mamouras}, \bibinfo{person}{Daniela Petrisan}, {and}
  \bibinfo{person}{Alexandra Silva}.} \bibinfo{year}{2015}\natexlab{}.
\newblock \showarticletitle{Nominal {K}leene Coalgebra}. In
  \bibinfo{booktitle}{\emph{Automata, Languages, and Programming, {ICALP}
  2015}} \emph{(\bibinfo{series}{lncs}, Vol.~\bibinfo{volume}{9135})}.
  \bibinfo{publisher}{springer}, \bibinfo{pages}{286--298}.
\newblock
\showISBNx{978-3-662-47665-9}
\urldef\tempurl%
\url{https://doi.org/10.1007/978-3-662-47666-6}
\showDOI{\tempurl}


\bibitem[Kwiatkowska et~al\mbox{.}(2011)]%
        {PRISM}
\bibfield{author}{\bibinfo{person}{Marta Kwiatkowska}, \bibinfo{person}{Gethin
  Norman}, {and} \bibinfo{person}{David Parker}.}
  \bibinfo{year}{2011}\natexlab{}.
\newblock \showarticletitle{PRISM 4.0: Verification of Probabilistic Real-Time
  Systems}. In \bibinfo{booktitle}{\emph{Computer Aided Verification}},
  \bibfield{editor}{\bibinfo{person}{Ganesh Gopalakrishnan} {and}
  \bibinfo{person}{Shaz Qadeer}} (Eds.). \bibinfo{publisher}{Springer Berlin
  Heidelberg}, \bibinfo{address}{Berlin, Heidelberg},
  \bibinfo{pages}{585--591}.
\newblock
\showISBNx{978-3-642-22110-1}


\bibitem[Larsen and Arne~Skou(1991)]%
        {LarsenS91}
\bibfield{author}{\bibinfo{person}{Kim~Guldstrand Larsen} {and}
  \bibinfo{person}{Arne Arne~Skou}.} \bibinfo{year}{1991}\natexlab{}.
\newblock \showarticletitle{Bisimulation through Probabilistic Testing}.
\newblock \bibinfo{journal}{\emph{Inform.\ Comput.}} \bibinfo{volume}{94},
  \bibinfo{number}{1} (\bibinfo{year}{1991}), \bibinfo{pages}{1--28}.
\newblock


\bibitem[May and Knight(2006)]%
        {MayKnight06}
\bibfield{author}{\bibinfo{person}{Jonathan May} {and} \bibinfo{person}{Kevin
  Knight}.} \bibinfo{year}{2006}\natexlab{}.
\newblock \showarticletitle{Tiburon: A Weighted Tree Automata Toolkit}. In
  \bibinfo{booktitle}{\emph{Implementation and Application of Automata}},
  \bibfield{editor}{\bibinfo{person}{Oscar~H. Ibarra} {and}
  \bibinfo{person}{Hsu-Chun Yen}} (Eds.). \bibinfo{publisher}{Springer Berlin
  Heidelberg}, \bibinfo{address}{Berlin, Heidelberg},
  \bibinfo{pages}{102--113}.
\newblock
\showISBNx{978-3-540-37214-1}


\bibitem[Milius et~al\mbox{.}(2015)]%
        {MiliusPS15}
\bibfield{author}{\bibinfo{person}{Stefan Milius}, \bibinfo{person}{Dirk
  Pattinson}, {and} \bibinfo{person}{Lutz Schr{\"{o}}der}.}
  \bibinfo{year}{2015}\natexlab{}.
\newblock \showarticletitle{Generic Trace Semantics and Graded Monads}. In
  \bibinfo{booktitle}{\emph{6th Conference on Algebra and Coalgebra in Computer
  Science, {CALCO} 2015}} \emph{(\bibinfo{series}{LIPIcs},
  Vol.~\bibinfo{volume}{35})}, \bibfield{editor}{\bibinfo{person}{Lawrence~S.
  Moss} {and} \bibinfo{person}{Pawel Sobocinski}} (Eds.).
  \bibinfo{publisher}{Schloss Dagstuhl - Leibniz-Zentrum f{\"{u}}r Informatik},
  \bibinfo{pages}{253--269}.
\newblock
\urldef\tempurl%
\url{https://doi.org/10.4230/LIPIcs.CALCO.2015.253}
\showDOI{\tempurl}


\bibitem[Milius et~al\mbox{.}(2016)]%
        {msw16}
\bibfield{author}{\bibinfo{person}{Stefan Milius}, \bibinfo{person}{Lutz
  Schr{\"o}der}, {and} \bibinfo{person}{Thorsten Wi{\ss}mann}.}
  \bibinfo{year}{2016}\natexlab{}.
\newblock \showarticletitle{Regular Behaviours with Names}.
\newblock \bibinfo{journal}{\emph{Applied Categorical Structures}}
  \bibinfo{volume}{24}, \bibinfo{number}{5} (\bibinfo{year}{2016}),
  \bibinfo{pages}{663--701}.
\newblock
\showISSN{1572-9095}
\urldef\tempurl%
\url{https://doi.org/10.1007/s10485-016-9457-8}
\showDOI{\tempurl}


\bibitem[Milner(1980)]%
        {Milner1980}
\bibfield{author}{\bibinfo{person}{Robin Milner}.}
  \bibinfo{year}{1980}\natexlab{}.
\newblock \bibinfo{booktitle}{\emph{A Calculus of Communicating Systems}}.
\newblock \bibinfo{publisher}{Springer Berlin Heidelberg}.
\newblock
\urldef\tempurl%
\url{https://doi.org/10.1007/3-540-10235-3}
\showDOI{\tempurl}


\bibitem[Moore(1956)]%
        {Moore}
\bibfield{author}{\bibinfo{person}{Edward~F. Moore}.}
  \bibinfo{year}{1956}\natexlab{}.
\newblock \bibinfo{booktitle}{\emph{Gedanken-Experiments on Sequential
  Machines}}.
\newblock \bibinfo{publisher}{Princeton University Press},
  \bibinfo{pages}{129--154}.
\newblock
\urldef\tempurl%
\url{https://doi.org/doi:10.1515/9781400882618-006}
\showDOI{\tempurl}


\bibitem[Paige and Tarjan(1987)]%
        {PaigeTarjan87}
\bibfield{author}{\bibinfo{person}{Robert Paige} {and}
  \bibinfo{person}{Robert~E.\ Tarjan}.} \bibinfo{year}{1987}\natexlab{}.
\newblock \showarticletitle{Three partition refinement algorithms}.
\newblock \bibinfo{journal}{\emph{SIAM J.~Comput.}} \bibinfo{volume}{16},
  \bibinfo{number}{6} (\bibinfo{year}{1987}), \bibinfo{pages}{973--989}.
\newblock


\bibitem[Parikh(1985)]%
        {Parikh1985}
\bibfield{author}{\bibinfo{person}{Rohit Parikh}.}
  \bibinfo{year}{1985}\natexlab{}.
\newblock \showarticletitle{The Logic of Games and its Applications}.
\newblock In \bibinfo{booktitle}{\emph{Topics in the Theory of Computation,
  Selected Papers of the International Conference on `Foundations of
  Computation Theory', {FCT} {\textquotesingle}83}}.
  \bibinfo{publisher}{Elsevier}, \bibinfo{pages}{111--139}.
\newblock
\urldef\tempurl%
\url{https://doi.org/10.1016/s0304-0208(08)73078-0}
\showDOI{\tempurl}


\bibitem[Pauly(2001)]%
        {Pauly2001}
\bibfield{author}{\bibinfo{person}{Marc Pauly}.}
  \bibinfo{year}{2001}\natexlab{}.
\newblock \emph{\bibinfo{title}{Logic for Social Software}}.
\newblock \bibinfo{thesistype}{Ph.\,D. Dissertation}.
\newblock
\urldef\tempurl%
\url{https://dare.uva.nl/search?identifier=9ad66ec5-063d-4673-8563-91369d0af7aa}
\showURL{%
\tempurl}


\bibitem[Peleg(1987)]%
        {Peleg87}
\bibfield{author}{\bibinfo{person}{David Peleg}.}
  \bibinfo{year}{1987}\natexlab{}.
\newblock \showarticletitle{Concurrent Dynamic Logic}.
\newblock \bibinfo{journal}{\emph{J. ACM}} \bibinfo{volume}{34},
  \bibinfo{number}{2} (\bibinfo{date}{apr} \bibinfo{year}{1987}),
  \bibinfo{pages}{450–479}.
\newblock
\showISSN{0004-5411}
\urldef\tempurl%
\url{https://doi.org/10.1145/23005.23008}
\showDOI{\tempurl}


\bibitem[Schr\"oder et~al\mbox{.}(2017)]%
        {skmw17}
\bibfield{author}{\bibinfo{person}{Lutz Schr\"oder}, \bibinfo{person}{Dexter
  Kozen}, \bibinfo{person}{Stefan Milius}, {and} \bibinfo{person}{Thorsten
  Wi\ss{}mann}.} \bibinfo{year}{2017}\natexlab{}.
\newblock \showarticletitle{Nominal Automata with Name Binding}. In
  \bibinfo{booktitle}{\emph{FoSSaCS 2017}} \emph{(\bibinfo{series}{LNCS},
  Vol.~\bibinfo{volume}{10203})}, \bibfield{editor}{\bibinfo{person}{Javier
  Esparza} {and} \bibinfo{person}{Andrzej Murawski}} (Eds.).
  \bibinfo{publisher}{Springer}, \bibinfo{pages}{124--142}.
\newblock
\urldef\tempurl%
\url{https://doi.org/10.1007/978-3-662-54458-7_8}
\showDOI{\tempurl}


\bibitem[Silva et~al\mbox{.}(2013)]%
        {SilvaBBR13}
\bibfield{author}{\bibinfo{person}{Alexandra Silva}, \bibinfo{person}{Filippo
  Bonchi}, \bibinfo{person}{Marcello~M. Bonsangue}, {and} \bibinfo{person}{Jan
  J. M.~M. Rutten}.} \bibinfo{year}{2013}\natexlab{}.
\newblock \showarticletitle{Generalizing determinization from automata to
  coalgebras}.
\newblock \bibinfo{journal}{\emph{Log. Methods Comput. Sci.}}
  \bibinfo{volume}{9}, \bibinfo{number}{1} (\bibinfo{year}{2013}).
\newblock
\urldef\tempurl%
\url{https://doi.org/10.2168/LMCS-9(1:9)2013}
\showDOI{\tempurl}


\bibitem[Silva and Sokolova(2011)]%
        {SILVA2011291}
\bibfield{author}{\bibinfo{person}{Alexandra Silva} {and} \bibinfo{person}{Ana
  Sokolova}.} \bibinfo{year}{2011}\natexlab{}.
\newblock \showarticletitle{Sound and Complete Axiomatization of Trace
  Semantics for Probabilistic Systems}.
\newblock \bibinfo{journal}{\emph{Electronic Notes in Theoretical Computer
  Science}}  \bibinfo{volume}{276} (\bibinfo{year}{2011}),
  \bibinfo{pages}{291--311}.
\newblock
\showISSN{1571-0661}
\urldef\tempurl%
\url{https://doi.org/10.1016/j.entcs.2011.09.027}
\showDOI{\tempurl}
\newblock
\shownote{Twenty-seventh Conference on the Mathematical Foundations of
  Programming Semantics (MFPS XXVII)}.


\bibitem[Trnkov\'a(1969)]%
        {trnkova69}
\bibfield{author}{\bibinfo{person}{V\v{e}ra Trnkov\'a}.}
  \bibinfo{year}{1969}\natexlab{}.
\newblock \showarticletitle{Some properties of set functors}.
\newblock \bibinfo{journal}{\emph{Commentationes Mathematicae Universitatis
  Carolinae}} \bibinfo{volume}{10}, \bibinfo{number}{2} (\bibinfo{year}{1969}),
  \bibinfo{pages}{323--352}.
\newblock


\bibitem[Trnkov\'a(1971)]%
        {trnkova71}
\bibfield{author}{\bibinfo{person}{V\v{e}ra Trnkov\'a}.}
  \bibinfo{year}{1971}\natexlab{}.
\newblock \showarticletitle{On a descriptive classification of set functors
  {I}}.
\newblock \bibinfo{journal}{\emph{Commentationes Mathematicae Universitatis
  Carolinae}} \bibinfo{volume}{12}, \bibinfo{number}{1} (\bibinfo{year}{1971}),
  \bibinfo{pages}{143--174}.
\newblock


\bibitem[Valmari(2009)]%
        {Valmari09}
\bibfield{author}{\bibinfo{person}{Antti Valmari}.}
  \bibinfo{year}{2009}\natexlab{}.
\newblock \showarticletitle{Bisimilarity Minimization in {$\CO(m \log n)$}
  Time}. In \bibinfo{booktitle}{\emph{Applications and Theory of Petri Nets,
  {PETRI} {NETS} 2009}} \emph{(\bibinfo{series}{LNCS},
  Vol.~\bibinfo{volume}{5606})}. \bibinfo{publisher}{Springer},
  \bibinfo{pages}{123--142}.
\newblock
\showISBNx{978-3-642-02423-8}


\bibitem[Valmari(2010)]%
        {Valmari10}
\bibfield{author}{\bibinfo{person}{Antti Valmari}.}
  \bibinfo{year}{2010}\natexlab{}.
\newblock \showarticletitle{Simple Bisimilarity Minimization in O(m log n)
  Time}.
\newblock \bibinfo{journal}{\emph{Fundam. Informaticae}} \bibinfo{volume}{105},
  \bibinfo{number}{3} (\bibinfo{year}{2010}), \bibinfo{pages}{319--339}.
\newblock
\urldef\tempurl%
\url{https://doi.org/10.3233/FI-2010-369}
\showDOI{\tempurl}


\bibitem[Valmari and Franceschinis(2010)]%
        {ValmariF10}
\bibfield{author}{\bibinfo{person}{Antti Valmari} {and}
  \bibinfo{person}{Giuliana Franceschinis}.} \bibinfo{year}{2010}\natexlab{}.
\newblock \showarticletitle{Simple {$\CO(m\log n)$} Time {M}arkov Chain
  Lumping}. In \bibinfo{booktitle}{\emph{Tools and Algorithms for the
  Construction and Analysis of Systems, TACAS 2010}}
  \emph{(\bibinfo{series}{LNCS}, Vol.~\bibinfo{volume}{6015})}.
  \bibinfo{publisher}{Springer}, \bibinfo{pages}{38--52}.
\newblock


\bibitem[Valmari and Lehtinen(2008)]%
        {ValmariLehtinen08}
\bibfield{author}{\bibinfo{person}{Antti Valmari} {and} \bibinfo{person}{Petri
  Lehtinen}.} \bibinfo{year}{2008}\natexlab{}.
\newblock \showarticletitle{Efficient Minimization of DFAs with Partial
  Transition}. In \bibinfo{booktitle}{\emph{Theoretical Aspects of Computer
  Science, {STACS} 2008}} \emph{(\bibinfo{series}{LIPIcs},
  Vol.~\bibinfo{volume}{1})}. \bibinfo{publisher}{Schloss Dagstuhl --
  Leibniz-Zentrum für Informatik, Germany}, \bibinfo{pages}{645--656}.
\newblock


\bibitem[van Glabbeek(2001)]%
        {Glabbeek01}
\bibfield{author}{\bibinfo{person}{Rob~J. van Glabbeek}.}
  \bibinfo{year}{2001}\natexlab{}.
\newblock \showarticletitle{The Linear Time - Branching Time Spectrum {I}}.
\newblock In \bibinfo{booktitle}{\emph{Handbook of Process Algebra}},
  \bibfield{editor}{\bibinfo{person}{Jan~A. Bergstra}, \bibinfo{person}{Alban
  Ponse}, {and} \bibinfo{person}{Scott~A. Smolka}} (Eds.).
  \bibinfo{publisher}{North-Holland / Elsevier}, \bibinfo{pages}{3--99}.
\newblock
\urldef\tempurl%
\url{https://doi.org/10.1016/b978-044482830-9/50019-9}
\showDOI{\tempurl}


\bibitem[Winskel(1993)]%
        {Winskel93}
\bibfield{author}{\bibinfo{person}{Glynn Winskel}.}
  \bibinfo{year}{1993}\natexlab{}.
\newblock \bibinfo{booktitle}{\emph{The formal semantics of programming
  languages - an introduction}}.
\newblock \bibinfo{publisher}{{MIT} Press}.
\newblock
\showISBNx{978-0-262-23169-5}


\bibitem[Wi{\ss}mann et~al\mbox{.}(2021)]%
        {WissmannEA2021}
\bibfield{author}{\bibinfo{person}{Thorsten Wi{\ss}mann},
  \bibinfo{person}{Hans-Peter Deifel}, \bibinfo{person}{Stefan Milius}, {and}
  \bibinfo{person}{Lutz Schr\"{o}der}.} \bibinfo{year}{2021}\natexlab{}.
\newblock \showarticletitle{From generic partition refinement to weighted tree
  automata minimization}.
\newblock \bibinfo{journal}{\emph{Formal Aspects of Computing}}
  (\bibinfo{date}{March} \bibinfo{year}{2021}), \bibinfo{pages}{1--33}.
\newblock
\urldef\tempurl%
\url{https://doi.org/10.1007/s00165-020-00526-z}
\showDOI{\tempurl}


\bibitem[Wißmann(2023)]%
        {SuppSet23}
\bibfield{author}{\bibinfo{person}{Thorsten Wißmann}.}
  \bibinfo{year}{2023}\natexlab{}.
\newblock \showarticletitle{Supported Sets – A New Foundation For Nominal
  Sets And Automata}. In \bibinfo{booktitle}{\emph{Computer Science Logic
  ({CSL'23})}} \emph{(\bibinfo{series}{LIPIcs})}.
\newblock
\urldef\tempurl%
\url{http://arxiv.org/abs/2201.09825}
\showURL{%
\tempurl}
\newblock
\shownote{to appear}.


\bibitem[Wißmann et~al\mbox{.}(2020)]%
        {concurSpecialIssue}
\bibfield{author}{\bibinfo{person}{Thorsten Wißmann}, \bibinfo{person}{Ulrich
  Dorsch}, \bibinfo{person}{Stefan Milius}, {and} \bibinfo{person}{Lutz
  Schröder}.} \bibinfo{year}{2020}\natexlab{}.
\newblock \showarticletitle{Efficient and Modular Coalgebraic Partition
  Refinement}.
\newblock \bibinfo{journal}{\emph{{Logical Methods in Computer Science}}}
  \bibinfo{volume}{{16:1}} (\bibinfo{date}{January} \bibinfo{year}{2020}),
  \bibinfo{pages}{8:1--8:63}.
\newblock
\urldef\tempurl%
\url{https://doi.org/10.23638/LMCS-16(1:8)2020}
\showDOI{\tempurl}


\end{thebibliography}

\ifthenelse{\boolean{dropappendix}}{}{%
  \newpage
  \appendix%
  \ifthenelse{\boolean{proofsinappendix}}{%
    \newpage%
    \section{Omitted Proofs}
    \closeoutputstream{proofstream}
    \input{\jobname-proofs.out}
  }{%
  }
}

\end{document}
